\documentclass[%
 reprint,
 superscriptaddress,
bibnotes,
amsmath,amssymb,
aps, 
pra,
]{revtex4-2}
\usepackage{physics}
\usepackage{amsmath}
\usepackage{braket}
\usepackage{bbold}
\usepackage{relsize}
\usepackage[svgnames]{xcolor}
\usepackage[caption=false,subrefformat=parens,labelformat=parens]{subfig}
\usepackage{hyperref}
\hypersetup{
    colorlinks=true,
    citecolor=blue,
    linkcolor=blue,
    filecolor=blue,  
    urlcolor=DarkBlue,
    pdftitle={Optimal negative-partial-transpose based entanglement criteria},
    pdfpagemode=FullScreen,
    }
\usepackage{url}
\usepackage{todonotes}
\setuptodonotes{fancyline,size=\tiny,color=green!40,shadow}

\usepackage{array}

\usepackage{orcidlink}

\begin{document}
\title{
Optimizing the statistical power of negative-partial-transpose-based \\ entanglement tests}
\author{Lydia A. Kanari-Naish\,\orcidlink{0000-0002-8089-5726}}
 \affiliation{Quantum Measurement Lab, Blackett Laboratory, \href{https://ror.org/041kmwe10}{Imperial College London}, London SW7 2BW, United Kingdom}
 \author{Amaya Calvo-S\'{a}nchez\,\orcidlink{0009-0006-8017-8870}
 }
 \affiliation{Quantum Measurement Lab, Blackett Laboratory, \href{https://ror.org/041kmwe10}{Imperial College London}, London SW7 2BW, United Kingdom}
\author{Jack Clarke\,\orcidlink{0000-0001-8055-449X}}
 \affiliation{Quantum Measurement Lab, Blackett Laboratory, \href{https://ror.org/041kmwe10}{Imperial College London}, London SW7 2BW, United Kingdom}
 \affiliation{Department of Physics and Astronomy, \href{https://ror.org/02jx3x895}{University College London}, Gower Street, WC1E 6BT London, United Kingdom}

\author{Arjun Gupta\,\orcidlink{0009-0009-2994-9125}}
 \affiliation{Quantum Measurement Lab, Blackett Laboratory, \href{https://ror.org/041kmwe10}{Imperial College London}, London SW7 2BW, United Kingdom}
\author{Sofia Qvarfort\,\orcidlink{0000-0003-2281-1042}}
\affiliation{Quantum Measurement Lab, Blackett Laboratory, \href{https://ror.org/041kmwe10}{Imperial College London}, London SW7 2BW, United Kingdom}
\affiliation{\href{https://ror.org/03r06fs10}{Nordita}, \href{https://ror.org/026vcq606}{KTH Royal Institute of Technology and Stockholm University}, Hannes Alfv\'{e}ns v\"{a}g 12, SE-106 91 Stockholm, Sweden}
\affiliation{Department of Physics, \href{https://ror.org/05f0yaq80}{Stockholm University}, AlbaNova University Center, SE-106 91 Stockholm, Sweden}
\author{Michael R. Vanner\,\orcidlink{0000-0001-9816-5994}}
\email{m.vanner@imperial.ac.uk}
\homepage{www.qmeas.net}
\affiliation{Quantum Measurement Lab, Blackett Laboratory, \href{https://ror.org/041kmwe10}{Imperial College London}, London SW7 2BW, United Kingdom}

\date{\today} 

\begin{abstract}
A key requirement of any separable quantum state is that its density matrix has a non-negative partial transpose. For continuous bipartite quantum states, violation of this condition may be tested via the hierarchy of negative-partial-transpose (NPT) based entanglement criteria introduced by Shchukin and Vogel~[\href{https://doi.org/10.1103/PhysRevLett.95.230502}{Phys. Rev. Lett. \textbf{95}, 230502 (2005)}].  
However, even for a known state, a procedure to determine the optimal NPT-based criterion within the hierarchy has, thus far, remained outstanding.
Here, we introduce a theoretical framework and numerical toolbox \textit{NPyT} to compare criteria performance via the statistical power of hypothesis testing for a known input state to enable the optimal entanglement criterion to be determined.  Our framework takes into account environmental noise and finite number of measurements, and optimally allocates measurement resources.
To demonstrate the utility of our approach, we apply our statistical framework to prominent examples of Gaussian and non-Gaussian states, including the two-mode squeezed vacuum state, the quanta-subtracted two-mode squeezed vacuum state, and the two-mode Schr\"odinger-cat state.
We discuss the optimal criteria identified for these states and compare them to other commonly employed criteria. For example, we identify an entanglement test for the two-mode squeezed vacuum state that has high statistical power and outperforms Simon's criterion. This work illustrates that the non-Gaussianity of the entanglement test distribution increases with test matrix size, and highlights that smaller test matrices generally provide better performance.
Beyond bipartite inseparability, our framework can be extended to a wide variety of other matrix-based tests of nonclassicality and multi-mode entanglement.
\end{abstract}


\maketitle

\section{Introduction}
Quantum entanglement exhibits correlations that cannot be described classically and is now being utilized in a wide array of quantum technologies and tests of fundamental physics~\cite{horodecki2009quantum}. There is considerable interest to detect entanglement in both qubit and continuous-variable systems using multiple measurement approaches~\cite{guhne2009entanglement, brunner2014bell}. Specifically, detecting entanglement with 
entanglement witnesses or inequalities is powerful because it is more resource efficient compared with full state-tomography. 
For qubits, necessary and sufficient criteria for separability have been developed~\cite{peres1996separability, Horodecki1996} and more recently, research has explored the effect of imperfect measurement settings for steering~\cite{tavakoli2024quantum} and multipartite entanglement detection~\cite{cao2024genuine}.
In continuous-variable systems, entangled states can be broadly classified as either Gaussian or non-Gaussian. Gaussian states are fully characterized by the first and second moments of the field operators, whereas non-Gaussian states require a more complex statistical description.
Notably, entangled Gaussian states are a proven workhorse for experimental quantum science, finding several applications~\cite{Adesso2007,weedbrook2012gaussian}, and have been realized across a wide array of experiments ranging from optical two-mode squeezed vacuum states~\cite{ou1992realization,furusawa1998unconditional,eberle2013stable} to entangled massive mechanical oscillators~\cite{ockeloen2018stabilized,kotler2021direct}.
Beyond Gaussian states, non-Gaussian states enable a wide variety of quantum information processing protocols such as the enhancement of teleportation schemes~\cite{opatrny2000improvement,cochrane2002teleportation,olivares2003teleportation}, Bell tests of nonlocality~\cite{garcia2004proposal,dell2007continuous,brunner2014bell}, and universal quantum computation~\cite{lloyd1999quantum,gottesman2001encoding}.
Excitingly, non-Gaussian entanglement has now been observed across a broad range of experimental platforms including optical~\cite{Ourjoumtsev2006,jeong2014generation,morin2014remote,ra2020non} and microwave fields~\cite{Wang2016,gao2019entanglement,wang2022flying}, the motional state of pairs of atomic ions~\cite{moehring2007entanglement,jost2009entangled}, superconducting qubits~\cite{song201710,xu2022metrological}, and even in massive mechanical resonators approaching the macroscopic scale~\cite{lee2011entangling}. 

A number of methods have been proposed by which both Gaussian and non-Gaussian entanglement can be detected. A prominent such method utilizes the fact that a non-negative partial transposition is necessary for separability of bipartite states~\cite{Peres1996,horodecki1996necessary,Horodecki1996}.
Building upon this technique, Shchukin and Vogel introduced an elegant formalism with an infinite hierarchy of determinants constructed from moments of the state, where the existence of a negative determinant indicates that the state is entangled~\cite{Shchukin2005}. 
These negative-partial-transpose (NPT) based criteria derive from matrices constructed from these moments and an individual criterion, when violated, provides a sufficient (but not necessary) condition for entanglement~\footnote{{N}ote that there exist entangled states which have a non-negative partial transpose and all of these are bound entangled}.
A number of well-known entanglement criteria~\cite{tan1999confirming,Mancini2002, Agarwal2005, Raymer2003} including the Duan inequality~\cite{Duan2000}, Simon's criterion~\cite{Simon2000}, and the Hillery--Zubairy criteria~\cite{hillery2006entanglement} have been shown to be individual criteria within the Shchukin--Vogel hierarchy~\cite{Adesso2007}.

This Shchukin--Vogel hierarchy provides an extensive means to establish entanglement; however, for a given continuous-variable bipartite state, until the present work, there was no systematic method for identifying the optimal entanglement criterion. For example, to detect the entanglement of Gaussian states, using up to second-order moments is necessary and sufficient~\cite{Simon2000}; however, despite this apparent simplicity, there are 31 NPT criteria to assess statistically.
In addition, due to measurement sampling errors and the influence of environmental interactions, experimental uncertainty propagates into these entanglement criteria. Such uncertainty decreases the probability of correctly identifying entanglement when entanglement is present, i.e. the statistical power of the entanglement tests, and therefore plays a decisive role in the detection of entanglement.
The optimal entanglement test will then depend on the specific state under consideration, its interaction with the environment, as well as the method by which the moments are measured. Furthermore, in contrast to Gaussian entanglement, NPT criteria based on second-order moments may fail to detect entanglement in non-Gaussian states~\cite{agarwal2005inseparability}, being sufficient rather than necessary and sufficient, in which case higher-order NPT criteria may need to be employed~\cite{Dell2006,Zhang2021}. 

In this article, we develop a statistical framework to identify the optimal NPT-based entanglement criterion for any bipartite continuous-variable state. The framework takes into account finite number of measurements and open-system dynamics including loss to determine the statistical power in rejecting the no-entanglement hypothesis for a given input state.
To perform this hypothesis test, we first optimally allocate measurement resources utilizing an analytic approximation of the estimator uncertainty, and we then numerically generate the non-Gaussian distribution of the test statistic using a Monte Carlo method to obtain the statistical power.
We implement this as the openly available \textit{\textbf{N}PT test optimization \textbf{Py}thon \textbf{T}oolbox} (\textit{NPyT})~\cite{lydia_code}.
Using this toolbox, we observe that lower-dimensional tests generally outperform higher-dimensional tests as a result of the test distributions becoming more heavy-tailed with increasing dimension. More specifically, we demonstrate the utility of \textit{NPyT} by applying it to (i) two-mode squeezed vacuum states, (ii) quanta-subtracted two-mode squeezed vacuum states, and (iii) two-mode Schr{\"o}dinger-cat states.
As an illustrative example, for the two-mode squeezed vacuum, we identify a lower dimensional criterion that has higher statistical power than other more commonly used criteria such as Simon's criterion or the Duan inequality. We anticipate that this framework will be useful for a broad range of further studies and experiments in order to select the best possible measurement set to observe non-classicality.

\section{NPT entanglement criteria}
\label{sec:entanglement_criteria}
\subsection{Overview}
\label{sec:overview_entanglement}
Before discussing our framework, we provide a brief overview of the entanglement criteria introduced by Shchukin and Vogel in Ref.~\cite{Shchukin2005}. 
There, an infinite hierarchy of inseparability criteria are introduced, which, in its entirety, provides a necessary and sufficient means to establish NPT of a continuous-variable bipartite quantum state. Each individual criterion within the hierarchy corresponds to an inequality, which constitutes a sufficient, but not necessary, entanglement condition.
The inequalities are constructed by taking the determinants of Hermitian matrices composed of expectation values of products of annihilation and creation operators. These expectation values are henceforth referred to as moments and the matrices are constructed according to specific rules, detailed below. The negativity of the determinant indicates NPT and thus entanglement.

We first consider an infinite-dimensional square matrix $\mathbf{M}$ comprising moments with respect to a bipartite state $\hat{\rho}_\mathcal{AB}$. The element of $\mathbf{M}$ with indices $i$ and $j$ is the moment
\begin{equation}
\label{eq:super_matrix_ij}
    \mathbf{M}_{ij}=\braket{\hat{a}^{\dag q_i} \hat{a}^{p_i}\hat{a}^{\dag n_j}\hat{a}^{m_j}\hat{b}^{\dag l_j}\hat{b}^{k_j}\hat{b}^{\dag r_i}\hat{b}^{s_i}}.
\end{equation}
Here, $\hat{a}$ ($\hat{a}^{\dag}$) and $\hat{b}$ ($\hat{b}^{\dag}$) are the annihilation (creation) operators associated with subsystem $\mathcal{A}$ and $\mathcal{B}$, respectively. The angled brackets indicate the quantum expectation value with respect to the state $\hat{\rho}_\mathcal{AB}$ and each field operator is raised to a power that depends on the matrix-element indices.
The mapping between the element indices and the powers is given in Appendix~\ref{app:entanglement_criteria}, which we succinctly describe here for convenience.
The $i^{\mathrm{th}}$ index is equal to the ordinal number of the ordered, but infinite, list of multi-indices $(p_i,q_i,r_i,s_i)$ and, likewise, the $j^{\mathrm{th}}$ index corresponds to the ordinal number of the multi-indices $(n_j,m_j,k_j,l_j)$. The ordering of the multi-indices is arbitrary but once a convention is picked it must be kept consistent between the two sets of multi-indices, such that $(p_N, q_N, r_N, s_N)=(n_N, m_N,k_N, l_N)$ for all positive integer values of $N$. The precise ordering we use in this work is specified in Table~\ref{tab:multi_index} in Appendix~\ref{app:entanglement_criteria} and the first 15 rows and columns of $\mathbf{M}$ are given in Eq.~\eqref{eq:super_matrix}. As an example, we list the first few here; $i=1$ corresponds to $(p_1,q_1,r_1,s_1)=(0,0,0,0)$, $i=2$ corresponds to $(p_2,q_2,r_2,s_2)=(1,0,0,0)$, $i=3$ corresponds to $(p_3,q_3,r_3,s_3)=(0,1,0,0)$, and so on. The ordering for the $j^{\mathrm{th}}$ index follows the same convention.

We now consider Hermitian submatrices of $\mathbf{M}$ with determinants that are the NPT criteria. As a first example, one may consider the leading principal minors $\mathrm{det}[\mathbf{M }_{N}]$, which are obtained by taking the determinant of the first $N$ rows and $N$ columns of $\mathbf{M}$. By application of Sylvester's criterion, it can be shown that the partial transposition of $\hat{\rho}_\mathcal{AB}$ is positive if and only if $\mathrm{det}[\mathbf{M}_{N}]>0$ for all $N$. Thus, if $\mathrm{det}[\mathbf{M}_{{N}}]<0$ for any $N$, then the state has a negative partial transpose and is inseparable. Note that use of all the leading principal minors determines the partial transposition of $\hat{\rho}_\mathcal{AB}$ is positive definite, not positive semi-definite, and is thus a restricted class of entanglement tests. To broaden the range of tests, one can consider a Hermitian submatrix $\mathbf{A}$, which is constructed by deleting rows and columns of $\mathbf{M}$ with the same index, i.e. a principal minor. If $\mathrm{det}[\mathbf{A}]\geq0$ for all $\mathbf{A}$, then the partial transposition of $\hat{\rho}_\mathcal{AB}$ is positive semi-definite. Demonstrating $\mathrm{det}[\mathbf{A}]<0$ for any $\mathbf{A}$ indicates NPT and is a sufficient criterion for entanglement. This condition  will be used extensively in our analysis.
%


This work addresses two challenges that arise when testing for entanglement utilizing these NPT criteria. Firstly, finding effective entanglement tests for a given state, and then establishing the statistical power of entanglement verification in the presence of experimental uncertainties. The former challenge arises even in the absence of experimental uncertainties because, in general, not all inequalities reveal entanglement. It should also be emphasized that the magnitude of the negativity is not significant and should not be interpreted as a monotonic measure of entanglement. Thus, it is crucial to consider the statistical significance of the determinant since in an experiment each moment will have an unavoidable uncertainty. In particular, a negative determinant with an uncertainty that is greater than its magnitude is clearly not a conclusive entanglement test. Therefore, the statistical distribution of the determinant must be considered when searching for the optimal entanglement test.

\subsection{Parameterizing determinants}
\label{sec:characterizing}
For Gaussian states, the first 5 rows and columns of $\mathbf{M}$ [cf. Eq.~\eqref{eq:super_matrix}] yield the necessary and sufficient entanglement criterion $\mathrm{det}[\mathbf{M}_5]<0$, which is equivalent to Simon's criterion~\cite{Simon2000}. Beyond this criterion, finding suitable NPT tests for non-Gaussian states can be challenging since higher-order moments may need to be evaluated. 

To enable a systematic search, note that since $\mathbf{A}$ is generated by deleting rows and columns from $\mathbf{M}$ in a pairwise way, $\mathbf{A}$ can be fully described by the indices of the rows and columns that remain after this deletion---for examples, see Tables~\ref{tab:TMSV}, \ref{tab:TMSV_sub}, and \ref{tab:TMSC} of Appendix~\ref{app:entanglement_criteria}. An important parameter in our characterization of $\mathbf{A}$ is the number of rows and columns $d$, i.e. $\mathrm{dim}[\mathbf{A}]=d \times d$.
For practical tests, $d$ should be kept sufficiently small to limit the number of moments that must be measured. Consequently, we anticipate 
that the overall error on our estimate of $\mathrm{det}[\mathbf{A}]$ increases as $d$ increases. 
Furthermore, it is favourable to keep the order of the moments, which appear in $\mathbf{A}$, as low as possible, as measuring higher-order moments is more experimentally challenging~\cite{ShchukinNonClass, Kanari2022}. Defining the order of the moment in Eq.~\eqref{eq:super_matrix_ij} as $q_i+p_i+n_j+m_j+l_j+k_j+r_i+s_i$, we therefore identify the parameter $n$ as the order of the highest-order moment in $\mathbf{A}$, i.e.
\begin{equation}
    n=\mathrm{max}[q_i+p_i+n_j+m_j+l_j+k_j+r_i+s_i].
\end{equation}
In Section~\ref{sec:simulations} we describe our method to search through all $\mathbf{A}$ to detect entanglement for a specific state, given constraints on $d$ and $n$. 

We illustrate the above ideas with the following example. It has been noted~\cite{Shchukin2005,Miranowicz2009,Kanari2022} that the following determinant 
\begin{equation}
\label{eq:S3_matrix}
S_{\mathrm{III}}=\begin{vmatrix}
1 & \langle{\hat{b}^\dag\rangle} & \langle{\hat{a}\hat{b}^\dag\rangle} \\
\langle{\hat{b}\rangle} & \langle{\hat{b}^\dag \hat{b}\rangle} & \langle{\hat{a} \hat{b}^\dag \hat{b}\rangle}\\
\langle{\hat{a}^\dag \hat{b}\rangle} & \langle{\hat{a}^\dag \hat{b}^\dag \hat{b}\rangle} & \langle{\hat{a}^\dag \hat{a} \hat{b}^\dag \hat{b}\rangle}
\end{vmatrix},
\end{equation}
is negative for any canonical odd two-mode Schr{\"o}dinger-cat state $\ket{\psi}\propto(\ket{\alpha}\ket{\beta}-\ket{-\alpha}\ket{-\beta}$), where $\alpha$ and $\beta$ are non-zero complex amplitudes.
Using the multi-index ordering in Table \ref{tab:multi_index}, the submatrix $\mathbf{A}$ that produces the determinant $S_{\mathrm{III}}$ is generated by keeping the first, fifth, and thirteenth rows and columns of $\mathbf{M}$. Here, the highest-order moment in the submatrix is $\braket{\hat{a}^\dag \hat{a}\hat{b}^\dag \hat{b}}$ so $n=4$ and the dimension of the submatrix is given by $d=3$. Here, the subscript label $\mathrm{III}$ is for naming purposes only.

\subsection{Local operations and rotational invariance of NPT criteria}
\label{sec:invariance}

Notably, the NPT tests described here are generally not invariant under local operations, such as displacements. This lack of invariance may seem surprising since entanglement cannot be generated by local operations, nor can the entanglement entropy be enhanced by local unitaries~\cite{Vedral1997}. However, we would like to remind our readers that NPT-based tests are a sufficient but not necessary test of entanglement, and the negativity of $\mathrm{det}[\mathbf{A}]$ is also not a monotonic measure of entanglement.

One exception to this behaviour is NPT criteria are invariant under local rotations, which we prove in Appendix~\ref{app:invariance}. 

\section{Framework for selecting optimal NPT criteria}
\label{sec:error_propagation}
In this section, we introduce and describe our statistical framework. We derive an analytic expression for the uncertainty on NPT criteria given by the standard deviation of the determinant estimator, which arises due to random sampling error associated with finite measurement statistics. 
Importantly, we also discuss how to optimally allocate measurement resources and thus minimize this overall uncertainty. Finally, we describe a hypothesis test and determine its statistical power to conclude entanglement for a given NPT criterion, which enables the optimal NPT criterion to be selected.

\subsection{Notation and statistics}
To establish our notation, we first briefly review the statistics of measuring independent and identically distributed random variables. Consider a sample of $M$ such variables which are obtained by sampling from an underlying distribution. Each measurement outcome $x_i$ has mean $E[x_i]=\mu$ and variance $E[x_i^2]-E[x_i]^2=\sigma^2=\textrm{Var}(x_i)$. To estimate $\mu$ we use the sample mean estimator $\bar{X} = (\sum_{i=1}^Mx_i)/M$. The error on $\bar{X}$ is quantified by the standard deviation of $\bar{X}$ given by $\sigma[\bar{X}]=\sqrt{E[\bar{X}-E[X]]^2} = \sqrt{\textrm{Var}(\bar{X})}$.
The expectation value of the sample mean is the same as the population mean, $E[\bar{X}]=\mu$. However, the standard deviation of the sample mean is dependent on $M$ as $\sigma[\bar{X}]=\sigma/\sqrt{M}$.

The same principles are applicable for a quantum observable $\hat{B}$. We use hats to denote quantum operators, and a bar to denote sample mean estimators. It follows that the quantum expectation value $\braket{\hat{B}}$ is equivalent to the estimator $\bar{B}$ in the limit of infinite independent measurements, where the quantum state is recreated between measurements. We denote the variance of the quantum operator as $\mathrm{Var}(\hat{B})$, and the standard deviation as $\sigma[\hat{B}]=\sqrt{\mathrm{Var}(\hat B)}$.

\subsection{Determinant estimator}

We now use these concepts to formulate a method for estimating the value of the determinant of $\mathbf{A}$.
Crucially, we need our estimator to be an unbiased and consistent random variable. Unbiased estimators have expectation values that are equal to the true value. For example, the sample mean estimator $\bar{X}$ above has mean $E[\bar{X}] = \mu$, which is the true mean, making it an unbiased estimator. Consistent estimators have probability distributions which converge to the true value as the sample size grows to infinity. For example, $\bar{B}$ above is a consistent estimator of the quantum expectation value $\langle{\hat{B}}\rangle$. Both of these properties are desirable for our estimator such that experimental data can provide reliable estimates of the true value of the determinant.

We denote the matrix of sample means of the moments as $\bar{\mathbf{A}}$. To ensure that the estimator is unbiased we assume that the upper and lower triangular entries of $\bar{\mathbf{A}}$ are measured independently, as imposing Hermiticity on $\bar{\mathbf{A}}$ introduces bias. Following this, we define our estimator of $\mathrm{det}[\mathbf{A}]$ as $\mathrm{Re}\{\mathrm{det}[\bar{\mathbf{A}}]\}$, which obeys $E[\mathrm{Re}\{\mathrm{det}[\bar{\mathbf{A}}]\}]=\mathrm{det}[\mathbf{A}]$. Further details of this estimator are outlined in Appendix~\ref{app:error_prop}.

While the diagonal elements of the Hermitian matrix $\mathbf{A}$ are all real, off-diagonal elements of $\mathbf{A}$ may be complex. Thus, the off-diagonal operators cannot always be directly measured but can always be decomposed into two Hermitian components. We denote the element of $\mathbf{A}$ with indices $i$ and $j$ as ${A}_{ij}=\braket{\hat{O}_{ij}}$, where $\hat{O}_{ij}=\hat{a}^{\dag q_i} \hat{a}^{p_i}\hat{a}^{\dag n_j}\hat{a}^{m_j}\hat{b}^{\dag l_j}\hat{b}^{k_j}\hat{b}^{\dag r_i}\hat{b}^{s_i}$ and the sample mean estimate of $A_{ij}$ as $\bar{A}_{ij}$. We next write
\begin{equation}
\label{eq:Hermitian_decomposition}
    \hat{O}_{ij}=\hat{B}_{ij,0}+i\hat{B}_{ij,1} \, ,
\end{equation}
using the notation $\hat{B}_{ij,p}$ to refer to these two Hermitian operators where the label $p=\{0,1\}$ specifies each operator such that
\begin{equation}
\label{eq:B_{ij,p}}    \hat{B}_{ij,p}=\frac{(i)^p\big[\hat{O}_{ij}^\dag+(-1)^p\hat{O}_{ij}\big]}{2}~.
\end{equation}
Measurement of both $\hat{B}_{ij,0}$ and $\hat{B}_{ij,1}$ to determine their sample means $\bar{B}_{ij,p}$ then allows $\bar{A}_{ij}$ to be determined. The determinant estimator $\mathrm{Re}\{\mathrm{det}[\bar{\mathbf{A}}]\}$ can then be determined, which thus depends on the $2d^2-d$ variables $\bar{B}_{ij,p}$.

\subsection{Determinant estimator variance}
To analyse how measurement errors on $\bar{B}_{ij,p}$ propagate through the determinant estimator and calculate the standard error $\Delta\mathrm{Re}\{\mathrm{det}[\bar{\mathbf{A}}]\}$, one must consider the covariances between the sample means $\bar{B}_{ij,p}$. In general, quantum operators can have a non-zero covariance $\mathrm{Cov}(\hat{B}_{ij,p},\hat{B}_{kl,q})\neq 0$ and to capture the covariance between these quantum operators, joint measurements of $\hat{B}_{ij,p}$ and $\hat{B}_{kl,q}$ must be made. However, the covariance between the sample means of two operators is necessarily zero if they are obtained independently, i.e. $\mathrm{Cov}(\bar{B}_{ij,p},\bar{B}_{kl,q})=\delta_{ik}\delta_{jl}\delta_{pq} \mathrm{Var}(\bar{B}_{ij,p})$, where $\mathrm{Var}(\bar{B}_{ij,p})$ is the variance of the sample mean. This is because $\bar{B}_{ij,p}$ and $\bar{B}_{kl,q}$ are not quantum operators but instead are two independent random variables obtained from distinct experimental realizations of the quantum state.

If $M_{ij,p}$ measurements are allocated to the operator $\hat{B}_{ij,p}$, then the corresponding sample mean is
\begin{equation}
\label{eq:mean_empirical}
    \bar{B}_{ij,p}=\frac{1}{M_{ij,p}}\sum_{m}^{M_{ij,p}} b^{(m)}_{ij,p}~,
\end{equation}
where $b^{(m)}_{ij,p}$ is the $m^{\text{th}}$ measurement result from the $M_{ij,p}$ independent realizations of the state. 
Each measurement result is obtained from the same underlying probability distribution with mean $\langle\hat{B}_{ij,p}\rangle$ and variance $\mathrm{Var}(\hat{B}_{ij,p})$. Thus, the variance of the sample mean estimator is given by $\mathrm{Var}(\bar{B}_{ij,p})=\mathrm{Var}(\hat{B}_{ij,p})/M_{ij,p}$.

If $\mathrm{Var}(\hat{B}_{ij,p})$ is unknown it can be empirically approximated using the unbiased estimator
\begin{equation}
\label{eq:var_empirical}
    \mathrm{Var}(\hat{B}_{ij,p})\approx\sum_{m=1}^{M_{ij,p}}\frac{(\bar{B}_{ij,p}-b_{ij,p}^{(m)})^2}{M_{ij,p}-1}~.
\end{equation}
We reiterate here that, depending on the particular experimental measurement scheme, obtaining $b^{(m)}_{ij,p}$ may require a combination of measurements and post-processing, in which case $M_{ij,p}$ quantifies the overall size of the measurement set from which the statistic is obtained.

Using the aforementioned definitions, we may now calculate $\Delta\mathrm{Re}\{\mathrm{det}[\bar{\mathbf{A}}]\}$. Using the steps outlined in Appendix \ref{app:error_prop}, we first expand $\mathrm{Re}\{\mathrm{det}[\bar{\mathbf{A}}]\}$ to second order about the true values $\langle\hat{B}_{ij,p}\rangle$, and we then approximate the standard error squared of the estimator $(\Delta \mathrm{Re}\{\mathrm{det}[\bar{\mathbf{A}}]\})^2$ as

\begin{widetext}
\begin{align}\label{eq:delta_detA2_secorder_main}
&(\Delta\mathrm{Re}\{\mathrm{det}[\bar{\textbf{A}}]\})^2 = \sum_{i,j,p}\mathrm{Re}\{(-i)^p\mathrm{adj}[\mathbf{A}]_{ij}\}\frac{\mathrm{Var}(\hat{B}_{ij,p})}{M_{ij,p}}\\\nonumber
&+ \frac{1}{2}\frac{1}{(d-2)!}\sum_{\substack{i,j,k,l,p,q\\ i\neq k,j\neq l}}\sum_{\substack{i_3, i_4, ..., i_d \\ j_3, j_4, ..., j_d}}\varepsilon_{iki_3...i_d}\varepsilon_{jlj_3...j_d}\mathrm{Re}\{(i)^{p+q}A_{i_3j_3}...A_{i_dj_d}\}\frac{\mathrm{Var}(\hat{B}_{ij,p})}{M_{ij,p}}\frac{\mathrm{Var}(\hat{B}_{kl,q})}{M_{kl,q}}\nonumber\\&+\mathcal{O}\left(\left(\frac{\mathrm{Var}(\hat{B})}{M}\right)^3\right)\, ,\nonumber
\end{align}
\end{widetext}

\noindent which requires $M_{ij,p}$ to be sufficiently large such that $\mathrm{Var}(\hat {B}_{ij,p})/M_{ij,p}\ll1$ and we can thus neglect the cubic and higher terms to a good approximation. The subscripts on the cubic order have been dropped for brevity. In Eq.~\eqref{eq:delta_detA2_secorder_main},  $\mathrm{adj}[{\mathbf{A}}]$ is the adjugate of matrix $\mathbf{A}$~\footnote{{T}he adjugate of a matrix is the transpose of its cofactor matrix, and so, for non-singular matrices, this can be expressed as $\mathrm{adj}[\mathbf{A}]=\mathrm{det}[\mathbf{A}] \mathbf{A}^{-1}$.}, and we also make use of the Levi--Civita tensor in $d$ dimensions $\varepsilon_{i_1i_2...i_d}$ which is the totally anti-symmetric tensor. The standard error squared of the determinant estimator is expressed in terms of $M_{ij,p}$ and the true means $\langle \hat{B}_{ij,p}\rangle$ and variances $\mathrm{Var}(\hat{B}_{ij,p})$. This functional form is useful in Section~\ref{sec:simulations} where exact quantum expectation values are used to evaluate $\Delta \mathrm{Re}\{\mathrm{det}[\bar {\mathbf{A}}]\}$. We'd also like to highlight that the second-order terms are important because some first-order contributions may vanish if a given $\mathrm{adj}[\mathbf{A}]_{ij}=0$, even though the corresponding $\mathrm{Var}(\hat{B}_{ij,p})\neq0$. In a real experiment the quantity $\Delta \mathrm{Re}\{\mathrm{det}[\bar {\mathbf{A}}]\}$ can be empirically estimated using the measurement data set $\{b^{(m)}_{ij,p}\}$ alone.
The variances of sample means which appear in Eq.~\eqref{eq:delta_detA2_secorder_main} can be inferred from the empirical estimates of the quantum variance in Eq.~\eqref{eq:var_empirical}. 
Finally, we noted in Section~\ref{sec:invariance} that $\mathrm{det}[\mathbf{A}]$ is invariant when the quantum state undergoes local rotations, and we would like to note here that, similarly, the estimator $\textrm{Re}\{\textrm{det}[\bar{\mathbf{A}}]\}$ and its uncertainty $\Delta\textrm{Re}\{\mathrm{det}[\bar{\mathbf{A}}]\}$ are also invariant under such rotations.

\subsection{Minimizing error on an NPT criterion}
\label{sec:minimizing_error_on_NPT}
In the previous section, we introduced $M_{ij,p}$ as the number of measurements allocated for each Hermitian operator $\hat{B}_{ij,p}$. Since some operators will have a larger variance than others, they thus require more measurements to accurately obtain $\bar{B}_{ij,p}$. Here, we describe how to determine the optimal choice of $M_{ij,p}$ to minimize the overall standard deviation $\Delta\mathrm{Re} \{\mathrm{det}[\bar{\mathbf{A}}]\}$, for a given total measurement number $M_{\mathrm{tot}}$. The total number of measurements used to compute $\mathrm{Re}\{\mathrm{det}[\bar{\mathbf{A}}]\}$ can be expressed as
\begin{equation}
    M_{\mathrm{tot}}=\sum_{i}^d M_{ii,p=0}+2\sum_{\substack{i,j\\i<j}}^d \sum_{p=0}^1M_{ij,p}\, ,
\end{equation}
where the first term sums over the diagonal elements and the second term sums over the off-diagonal elements assuming $M_{ij,p}=M_{ji,p}$.
Using a Lagrange-multiplier method outlined in Appendix~\ref{app:minimizing_std}, we minimize $(\Delta \mathrm{Re}\{\mathrm{det}[\bar{\mathbf{A}}]\})^2$ defined in Eq.~\eqref{eq:delta_detA2_secorder_main} under the constraint of a fixed $M_{\mathrm{tot}}$. The optimization equations in Appendix~\ref{app:minimizing_std} are analytically intractable in general because they form a system of non-linear equations in $2d^2-d$ variables. Hence, we employ a numerical gradient descent method to obtain the optimal measurement allocation.
This allows us to best compare the performance of NPT criteria for the same number of measurement resources.

In general, this optimization leads to non-integer values for $M_{ij,p}$. We find a near-optimal integer set of $M_{ij,p}$ by first rounding the optimal solutions to the nearest integer, calculating the sum of these, and distributing the excess or deficit to satisfy the $M_{\rm{tot}}$ constraint.
This is done through a greedy redistribution algorithm: for a deficit of measurements,
unit increments are first made to the moment with the largest $M_{ij,p}$, then to the next largest in descending order until the deficit goes to 0. Similarly, if there is an excess of measurements, unit measurements are taken away first from the moment with the lowest $M_{ij,p}$, and then in ascending order. In both of these cases, we impose that the minimum number of measurements is $1$ (except for the identity which is not measured).

\subsection{Hypothesis testing and statistical power}
Following measurement-resource allocation, one may compute the statistical power of the hypothesis test where the statistic is $\mathrm{Re}\{\mathrm{det}[\bar{\mathbf{A}}]\}$. Namely, for a given NPT criterion set by the matrix of moments $\mathbf{A}$, the null hypothesis $H_0^{\mathbf{A}}$ is that no entanglement is observed and the alternative hypothesis $H_1^{\mathbf{A}}$ is that the state is entangled:
\begin{eqnarray}
    &H_0^{\mathbf{A}}:\quad{} \mathrm{det}[\mathbf{A}]\geq0 \quad{\text{(separable)}},\nonumber\\
    &H_1^{\mathbf{A}}:\quad{} \mathrm{det}[\mathbf{A}]<0 \quad{\text{(entangled)}}.
\end{eqnarray}
If we reject the no-entanglement hypothesis $H_{0}$, we must therefore accept $H_{1}$, i.e. that the state is entangled.
The statistical power of the test 
is defined as the probability of rejecting $H_0^{\mathbf{A}}$ when $H_1^{\mathbf{A}}$ is true, i.e. $P(\textrm{reject}\,H_0^{\mathbf{A}}|H_1^{\mathbf{A}})$. Assuming that the true value $\textrm{det}[\mathbf{A}]<0$, this probability is given as $P(\textrm{Re}\{\textrm{det}[\bar{\mathbf{A}}]\}<0)$. In Section~\ref{sec:overview_framework} below, we discuss how we generate the probability distribution of $\textrm{Re}\{\textrm{det}[\bar{\mathbf{A}}]\}$ via a Monte Carlo method.

The statistical power may then be calculated for a range of suitable NPT criteria and the one with the highest statistical power is identified as the optimal NPT criterion.

\section{Open quantum systems \label{sec:decoherence}}
All quantum systems interact with their environment, which affects the state via dissipation, thermalization, and other decoherence processes. Such interactions degrade any entanglement between the two subsystems, and thus make the null hypothesis more likely.
Namely, interactions with the environment between state generation and verification change the values of both the test statistic and its uncertainty, and therefore affect the confidence of NPT tests.

There are many unique noise models that apply to different open quantum systems. To demonstrate how the propagation of errors affects NPT criteria, here we consider a simple and widely used model to capture interactions of the state $\hat{\rho}_\mathcal{AB}$ with its environment.
Namely, we consider a beam-splitter model between system and environment, which is applicable to a wide range of continuous-variable systems~\cite{Leonhardt1997}. Considering the first subsystem $\mathcal{A}$, the field operator $\hat{a}$ after the beam-splitter interaction with the environment becomes
\begin{equation}
\label{eq:beam_splitter}
\hat{a}=\hat{a}_\mathrm{0}\sqrt{\eta}+\hat{a}_{E}\sqrt{1-\eta}~,
\end{equation}
where $\hat{a}_{\mathrm{0}}$ is the initial mode of interest, $\eta$ is the beam-splitter parameter, which equals $1$ in the absence of open-system dynamics, and $\hat{a}_{E}$ is the annihilation operator for the environmental mode. The environment is uncorrelated with $\hat{a}_{\mathrm{0}}$, and is in a thermal state such that
\begin{subequations}
    \begin{align}
        \braket{\hat{a}^\dag_E\hat{a}_E}&=\bar{N},\label{eq:environment_corr_1}\\
            \braket{\hat{a}_E \hat{a}^\dag_E}&=\bar N+1\label{eq:environment_corr_2},
    \end{align}
\end{subequations}
where $\bar{N}$ is the mean thermal occupation number of the bath.
For optical systems, their environmental bath is well described by $\bar{N}=0$ as the system operates at very high frequencies, whereas other lower-frequency systems, such as mechanical oscillators, usually have $\bar{N} > 0$.
Eqs.~\eqref{eq:beam_splitter}, \eqref{eq:environment_corr_1}, and \eqref{eq:environment_corr_2} also apply to subsystem $\mathcal{B}$ when $\hat{a}$ is substituted for $\hat{b}$. For brevity, in this work we assume that the environmental modes, $\hat{a}_{E}$ and $\hat{b}_{E}$, are independent and that both subsystems couple to the environment with the same $\eta$ and $\bar{N}$. However, our framework may be readily adapted to relax these assumptions if necessary.

In order to experimentally determine the entries in $\mathbf{A}$, moments such as $\braket{\hat{a}^\dag\hat{a}\hat{b}^\dag\hat{b}}$ must be measured. To predict the value of this measurement in the presence of open-system dynamics, we can use Eq. \eqref{eq:beam_splitter}  to write these moments in terms of the beam-splitter parameter $\eta$, the initial moment $\braket{\hat{a}^\dag_0\hat{a}_0\hat{b}^\dag_0\hat{b}_0}$, and correlations between field operators of the environment. In calculating the latter correlations, we use identities such as
\begin{equation}\label{eq:environment_correlations}
    \braket{\hat{a}^{\dag p}_E \hat{a}_E^q}=\delta_{pq}p! \bar{N}^p,
\end{equation}
which is derived in Appendix~\ref{app:open_system}. Using Eqs.~\eqref{eq:delta_detA2_secorder_main}, \eqref{eq:beam_splitter}, and \eqref{eq:environment_correlations}, we see the dependence of the measured moments on $\eta$ and $\bar{N}$, and how this propagates into $\mathrm{det}[\mathbf{A}]$ and $\Delta\mathrm{Re}\{\mathrm{det}[\bar{\mathbf{A}}]\}$.

An open-system dynamics approach can be adopted for platforms such as optical fields inside cavities and mechanical oscillators, where the quantum Langevin equations are an appropriate description~\cite{meystre2021quantum}. Notably, one can switch from the beam-splitter model for loss to the quantum-Langevin approach by substituting $\sqrt{\eta}=\mathrm{e}^{-\kappa t}$, where $\kappa$ is the amplitude decay rate and $t$ is the time between state generation and verification. We note that this approach therefore assumes that all moments are measured at the same time $t$ and the equivalence between the two approaches is outlined in Appendix~\ref{app:open_system}.

\section{Implementation with examples}
\label{sec:simulations}
 
In this section, we apply our statistical framework described in Section~\ref{sec:error_propagation}---including the environmental interactions described in Section~\ref{sec:decoherence}---to test entanglement in three example states:
(i) the two-mode squeezed vacuum (TMSV) state, (ii) the quanta-subtracted TMSV state, and (iii) a two-mode Schr{\"o}dinger-cat state. 
We first discuss these example states, we then summarize the steps of our framework, and finally we find the optimal NPT tests with the statistical powers obtained.

\subsection{Gaussian and non-Gaussian states of interest}\label{sec:states_of_interest}
\subsubsection{The two-mode squeezed vacuum state}
Firstly, we consider a prominent Gaussian state---the two-mode squeezed vacuum (TMSV) state---which is a cornerstone of quantum optics owing to the state's experimental availability and ability to possess a high degree of continuous-variable entanglement~\cite{ou1992realization,braunstein2005quantum}. Such states are key resources for the realization of many quantum technologies and protocols including quantum optical computing~\cite{lloyd1999quantum}, quantum key distribution~\cite{grosshans2003quantum}, and surpassing the standard quantum limit~\cite{ou1997fundamental,anisimov2010quantum}. Mathematically, the TMSV state is defined as
\begin{equation}
\label{eq:TMSV}
    \ket{\Psi}_{\mathrm{TMSV}}= e^{\frac{1}{2}\left(\zeta^* \hat{a}\hat{b}-\zeta \hat{a}^\dag \hat{b}^\dag\right)}\ket{0,0},
\end{equation} 
where the complex squeezing parameter is $\zeta=|\zeta|e^{i \phi}$ and $\phi = \mathrm{arg}(\zeta)$. Applying the rotation ${\hat{U}}(\phi,0)$ to cancel the phase in the two-mode squeezer gives the following state 
\begin{equation}
    {\hat{U}}(\phi,0 )\ket{\Psi}_{\mathrm{TMSV}} = e^{\frac{1}{2}\left(|\zeta| \hat{a}\hat{b}-|\zeta| \hat{a}^\dag \hat{b}^\dag\right)}\ket{0,0}.
\end{equation}
We therefore consider $\zeta \in \mathbb{R}^{+}$, as our estimator $\textrm{Re}\{\mathrm{det}[\bar{\mathbf{A}}]\}$ is invariant under local rotations, see Section~\ref{sec:invariance}. Note that the definition of the squeezing parameter used here was chosen to be consistent with the QuTiP package~\cite{johansson2012qutip} and e.g. Ref.~\cite{weedbrook2012gaussian}, and may differ to other parts of the quantum optics literature, see e.g. Refs~\cite{braunstein2005quantum, birrittella2023unconditional}.

\subsubsection{The quanta-subtracted two-mode squeezed vacuum state}

Secondly, we consider quanta-subtraction operations to the TMSV state~\cite{hong1999statistical,opatrny2000improvement}. Subtraction and addition operations can be utilized to create highly non-Gaussian states, which possess nonclassical statistical properties~\cite{Barnett2018}. 
Experimentally, for single-mode optical states, such operations have been implemented to create `kitten' states~\cite{Ourjoumtsev2006,Neergaard2006} as well as single-photon-added states~\cite{zavatta2004quantum,zavatta2007experimental}.
More recently, phonon addition and subtraction operations have enabled the observation of non-Gaussian states of mechanical motion~\cite{Vanner2013, milburn2016nonclassical, Enzian2021, Enzian2021nonG, patel2021room, patil2022measuring, cryer2025enhanced, clarke2025theoretical} and the moments of such mechanical states may be measured for the present work utilizing pulsed optomechanics~\cite{vanner2011pulsed,vanner2013cooling,muhonen2019state, Kanari2022}. 
Notably, applying such subtraction and addition operations to the two-mode squeezed vacuum state can be used to enhance entanglement~\cite{ourjoumtsev2007increasing,adesso2009experimentally,zhang2010distillation,takahashi2010entanglement,Navarrete2012}. Up to a normalization, the quanta-subtracted TMSV state is defined as
\begin{equation}
\label{eq:subTMSV}
    \ket{\Psi}_{\mathrm{SUB}}\propto \hat{a}^l \hat{b}^k \ket{\Psi}_{\mathrm{TMSV}},
\end{equation}
which describes a TMSV state subject to $l$- and $k$-quanta subtractions on modes $\mathcal{A}$ and $\mathcal{B}$, respectively. As with the TMSV state, we also take $\zeta$ to be real and positive here due to the invariance of $\mathrm{det}[\mathbf{A}]$ under local rotations. 

\subsubsection{The two-mode Schr{\"o}dinger-cat state}

As a final example state, we consider the non-Gaussian two-mode Schr{\"o}dinger-cat state, which finds broad interest across a range of applications and fundamental studies including universal and fault-tolerant quantum computation~\cite{munro2000entangled,jeong2002efficient,sanders2012review}, quantum metrology~\cite{gilchrist2004Schrodinger,giovannetti2011advances,Joo2011}, quantum macroscopicity~\cite{LeeCW2011,Yadin2016,clarke2018growing}, and investigations of the quantum-to-classical transition~\cite{Bose1999, Marshall2003, Kleckner2008, kanari2021can}.
Additionally, schemes to create such two-mode cat states in mechanical systems have been proposed~\cite{akram2013entangled,Kanari2022} and methods to verify their non-Gaussian entanglement with different Shchukin--Vogel criteria have also been studied~\cite{Kanari2022}. The two-mode Schr{\"o}dinger-cat state may be given by
\begin{equation}
\label{eq:cat_state}
    \ket{\Psi}_{\mathrm{CAT}}\propto\big(\ket{\alpha}\ket{0} - \ket{0}\ket{\alpha}\big) \,,
\end{equation} 
where $\ket{\alpha}$ is a coherent state given by the action of the displacement operator $\hat D(\alpha)=\exp(\alpha \hat a^\dag-\alpha^*\hat a)$ on the vacuum~\footnote{{T}he canonical Schr{\"o}dinger-cat state for subsystems $\mathcal{A}$ and $\mathcal{B}$ is usually defined as $\ket{\Psi}_{\mathrm{CAT}}\propto\ket{\alpha'}_{\mathcal{A}}\ket{\beta'}_{\mathcal{B}}-\ket{-\alpha'}_{\mathcal{A}}\ket{-\beta'}_{\mathcal{B}}$. The state we consider has a similar structure as it can be brought into this form for $|\alpha'|=|\beta'|=|\alpha|/2$ using local displacement operators and rotations, i.e. $\hat D_{\mathcal{A}}(-\alpha')\hat D_{\mathcal{B}}(\beta')\hat U(\mathrm{arg(\alpha'/\alpha),\mathrm{arg(\beta'/\alpha)}+\pi})\big(\ket{\alpha}\ket{0}-\ket{0}\ket{\alpha}\big) \propto \ket{\alpha'}\ket{\beta'} - \ket{-\alpha'}\ket{-\beta'}$.}. 
In general $\alpha \in \mathbb{C}$, however, the rotational invariance of the NPT criteria allows us to consider only real $\alpha$. That is, with $\alpha=|\alpha|e^{i\theta}$ where $\theta=\mathrm{arg}(\alpha)$, rotating the state $\ket{\Psi}_{\mathrm{CAT}}$ in Eq.~\eqref{eq:cat_state} via ${\hat{U}}(\theta,\theta)$ gives a state proportional to $\ket{|\alpha|}\ket{0}-\ket{0}\ket{|\alpha|}$, which yields the same value of $\mathrm{det}[\mathbf{A}]$ as $\ket{\Psi}_{\mathrm{CAT}}$.\\

\subsection{Overview of framework}\label{sec:overview_framework}

For a given state and its corresponding open-system dynamics, determining the optimal NPT criteria is challenging due to the very large number of possible determinants. To address this challenge, we have developed \textit{NPyT} to determine the optimal NPT test for a finite $d$ and $n$. The following summarizes the main capabilities and the key steps:

\begin{enumerate}
    \item We perform a preliminary search to reduce the number of candidate matrices $\mathbf{A}$ assuming there is no coupling to the environment and no sampling errors. Only determinants $\mathrm{det}[\mathbf{A}]$ that are negative in the absence of such experimental imperfections can be confidently negative when such effects are included. 
    Within a specified parameter range, determinants that exhibit such negativity in the absence of these experimental uncertainties are kept, and the rest are discarded.


    \item We then investigate the effects of environmental interactions
    on this subset of determinants. The open-system dynamics are incorporated using Eqs.~\eqref{eq:beam_splitter} and \eqref{eq:environment_correlations}, using the moments calculated in step 1. The determinants $\mathrm{det}[\mathbf{A}]$ are then obtained as a function of the efficiency parameter $\eta$ and the thermal occupation of the environment $\bar{N}$.
    
    

    \item For a finite number of total measurements $M_{\mathrm{tot}}$, the error on the determinant strongly depends on how the measurements are distributed among the matrix elements. We quantify the error on the determinant using the standard deviation of our determinant estimator $\Delta\rm Re\{\rm det[\bar{\mathbf{A}}]\}$. We apply a numerical gradient descent algorithm in \textit{NPyT} to find the optimal measurement allocation for a given state, $\eta$, and $\bar{N}$, by minimizing $\Delta\rm Re\{\rm det[\bar{\mathbf{A}}]\}$ under the constraint of finite total measurements $M_{\mathrm{tot}}$. We then apply a greedy algorithm to obtain a set of integer measurement numbers using the values obtained through numerical optimization as discussed in Section~\ref{sec:minimizing_error_on_NPT}. 
    

     \item For each NPT criterion characterised by a matrix of moments $\mathbf{A}$, we generate the probability distribution of $\textrm{Re}\{\textrm{det}[\bar{\mathbf{A}}]\}$ via a Monte Carlo method implemented in \textit{NPyT}. A Monte Carlo approach is used due to the nonlinear nature of the determinant. We model the distributions of the estimators of the real and imaginary parts $\bar{B}_{ij,0}$ and $\bar{B}_{ij,1}$ of the non-Hermitian elements as Gaussian distributions, since these elements are indefinitely signed, and we model $\bar{B}_{ij,0}$ of the Hermitian elements as gamma distributions, since these elements are positive-semidefinite. The gamma distribution captures the positive skewness of these elements and tends towards a Gaussian in the limit of large measurement number. Both the Gaussian and gamma distributions require two parameters, the mean and variance, which we obtain from the state and the measurement number. We then use a Monte Carlo method to sample from each matrix-element distribution to obtain the distribution of the determinant. This step is further discussed in Appendix~\ref{app:monte_carlo}.

    \item 
    We next numerically validate if the non-Hermitian matrix-element distributions converge sufficiently to be accurately approximated by a Gaussian with the assigned measurement number. (Such a check is not performed for the positive semi-definite gamma distributions, as these distributions capture such statistics well even for small measurement number.) 
    To do this we compute the excess kurtosis. If this quantity is sufficiently small, the approximation is valid, otherwise a larger $M_{\mathrm{tot}}$ is needed.
    
    \item Finally, we quantify the success of a determinant by calculating the statistical power of the hypothesis test in rejecting the no-entanglement hypothesis. 
    For a given state and set of system parameters, the $\mathbf{A}$-dependent hypothesis test with the highest statistical power identifies the optimal NPT criterion.
\end{enumerate}

\begin{figure*}
\centering
\includegraphics[width=\textwidth]{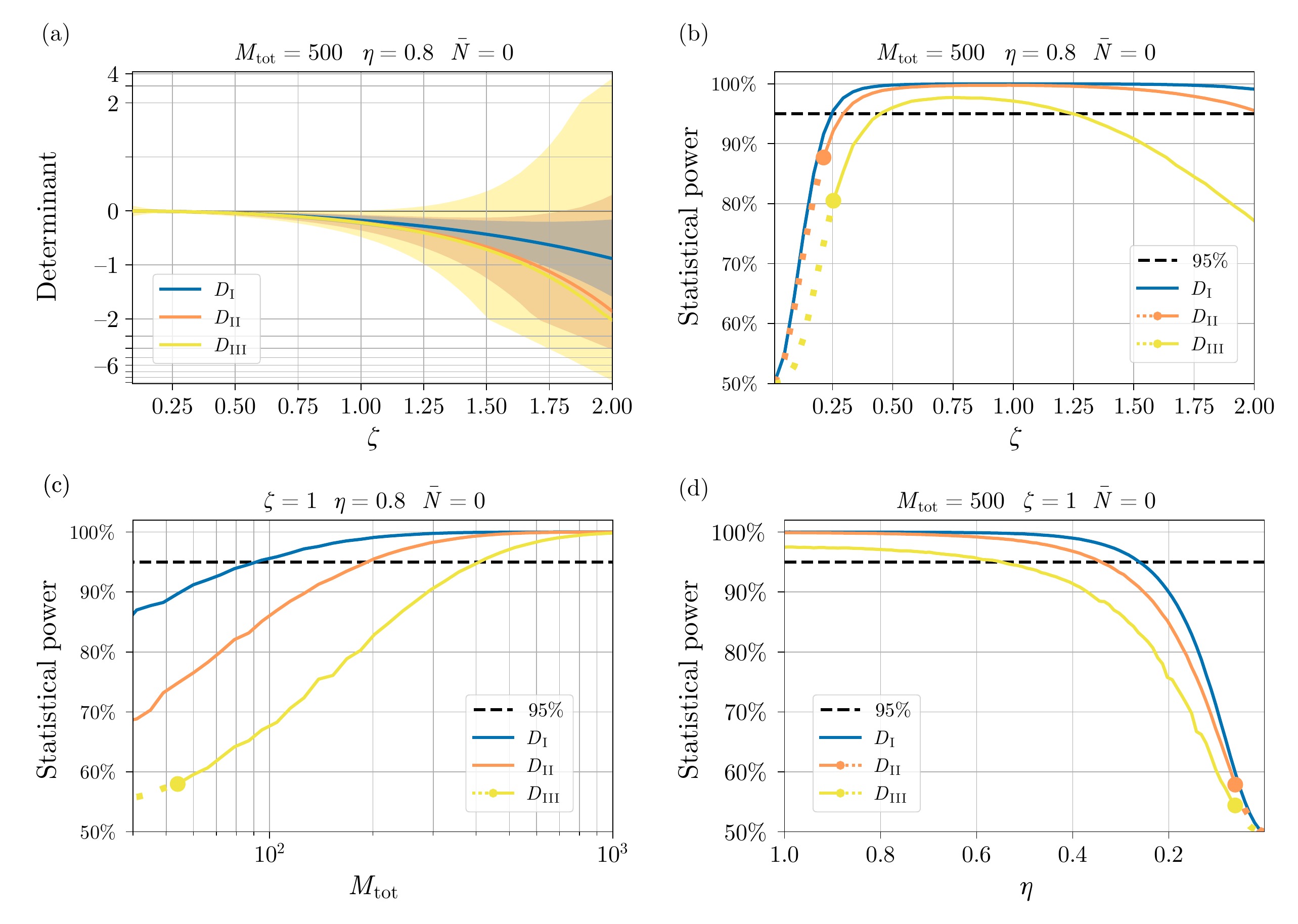}

\caption{NPT entanglement criteria for the two-mode squeezed vacuum state $\ket{\Psi}_{\mathrm{TMSV}}$ explored using \textit{NPyT}. Here, we plot the unique NPT criteria $\mathrm{det}[\mathbf{A}]$ that are capable of detecting entanglement within the $d\leq5$ and $n\leq 2$ subset that captures all Gaussian states. We explore their behavior and statistical power as functions of the squeezing parameter $\zeta$, the loss parameter $\eta$ for a bath thermal occupation $\bar{N}=0$, and the total number of measurements allocated to each determinant $M_{\mathrm{tot}}$. 
The parameters used for each plot are given as plot headers.
The dotted lines indicate parameter regions where an excess kurtosis of 4 is exceeded for at least one matrix element. 
(a) The unique determinants $D_{\mathrm{I-III}}$, which each exhibit negativity in the range $0\leq\zeta\leq2$. For this state, $D_{\mathrm{III}}$ is equivalent to Simon's criterion. The semi-transparent shaded regions denote the intervals from the $2.5$th to $97.5$th percentile of the non-Gaussian determinant distributions. 
(b) Statistical power with two-mode squeezing parameter $\zeta$. We find that for this parameter range, $D_\mathrm{I}$ is optimal for all $\zeta$ considered. For small squeezing, it is also the only determinant for which the matrix elements obey our choice of excess kurtosis threshold.
(c) Statistical power with $M_{\mathrm{tot}}$ for the range $10\leq M_{\mathrm{tot}}\leq10^3$. For these parameters, $D_\mathrm{I}$ requires the fewest number of measurements to reach a given statistical power, while $D_\mathrm{III}$ requires the most. 
(d) Statistical power with loss parameter $\eta$. For a lossless system with $\eta=1$, $D_{\mathrm{I-III}}$ have confidence levels above $95$\%, but as interactions with the environment increase these determinants degrade. Notably, $D_\mathrm{I}$ maintains the highest statistical power as loss is increased.}
\label{fig:results_TMSV}
\end{figure*}

\subsection{Results}\label{sec:results_section}

\subsubsection{Two-mode squeezed vacuum state}\label{sec:TMSV}

Here, we present the results for the TMSV state $\ket{\Psi}_{\mathrm{TMSV}}$, given in Eq.~\eqref{eq:TMSV}, as a prominent example of a Gaussian state. We consider the $31$ matrices $\mathbf{A}$ that satisfy ${d\leq5}$ and ${n\leq2}$, which capture all Gaussian states. Using the steps outlined in Section~\ref{sec:overview_framework} we identify eight determinants capable of detecting entanglement across the entire parameter range explored in Fig.~\ref{fig:results_TMSV}. These eight determinants are labelled as $D_{\mathrm{I-VIII}}$ and their corresponding matrices are shown explicitly in Table~\ref{tab:TMSV}. 
However, as may be seen from Table~\ref{tab:TMSV}, $D_{\mathrm{I}}=D_{\mathrm{IV}}$, $D_{\mathrm{II}}=D_{\mathrm{V}}$, and $D_{\mathrm{III}}=D_{\mathrm{VI}}$ for any state with $\braket{\hat{a}}=\braket{\hat{b}}=0$, such as the TMSV.
In addition, $D_\mathrm{VII}$ and $D_{\mathrm{VIII}}$ are also not unique owing to the symmetry of the TMSV state when the subsystems $\mathcal{A}$ and $\mathcal{B}$ are swapped. Namely, by swapping $\hat{a}$ ($\hat{a}^\dag$) with $\hat{b}$ ($\hat{b}^\dag$), one finds that $D_{\mathrm{II}}=D_{\mathrm{VII}}$ and $D_\mathrm{V}=D_{\mathrm{VIII}}$. We are then left with three unique determinants $D_\mathrm{I-III}$ that are capable of detecting the entanglement of this state. We also note that, for this state where $\braket{\hat{a}}=\braket{\hat{b}}=0$, $D_\mathrm{III}$ corresponds to Simon's criterion~\cite{Simon2000} and thus our work here provides a comparison of this well-known criterion to all other relevant Gaussian criteria.

In Fig.~\hyperref[fig:results_TMSV]{\ref*{fig:results_TMSV}(a)}, we plot the three unique determinants $D_\mathrm{I-III}$ with the two-mode squeezing parameter $\zeta$. Here, we assume an optical implementation with an optical loss of $\eta=0.8$ where the optical environment is well described by the vacuum state ($\bar{N}=0$). To demonstrate the uncertainty that arises due to limited measurement resources, the total number of measurements is chosen to be $M_{\mathrm{tot}}=500$. Fig.~\hyperref[fig:results_TMSV]{\ref*{fig:results_TMSV}(a)} also shows
a semi-transparent region representing a probability interval between the 2.5th and 97.5th percentiles, determined numerically. 
For this state, this uncertainty is parameterized by the total number of measurements $M_{\mathrm{tot}}$ and the optical loss parameter $\eta$ since we are setting $\bar{N}=0$. In the absence of such environmental interactions and sampling errors, i.e. $\Delta \textrm{Re}\{\mathrm{det}[\bar{\mathbf{A}}]\}=0$, the conditions $D_{\mathrm{I-II}}<0$ are sufficient criteria for entanglement for this Gaussian state and $D_{\mathrm{III}}<0$ is necessary and sufficient via Simon's criterion. However, when environmental interactions and sampling errors are included, these tests alone have sub-unity statistical power in determining entanglement due to the non-zero values of $\Delta \textrm{Re}\{\mathrm{det}[\bar{\mathbf{A}}]\}$. From Fig.~\hyperref[fig:results_TMSV]{\ref*{fig:results_TMSV}a}, this effect is especially apparent for the largest values of $\zeta$ where $\Delta \mathrm{Re}\{\mathrm{det}[\mathbf{A}]\}$ is similar in magnitude to $|\mathrm{det}[\mathbf{A}]|$ and is most prominent for Simon's criterion $D_{\mathrm{III}}$.

The statistical power of the NPT criteria in detecting entanglement is displayed in Fig.~\hyperref[fig:results_TMSV]{\ref*{fig:results_TMSV}(b)}, where we have plotted the statistical power with $\zeta$. 
To aid comparison, we have included the 95\% statistical power as a black dashed line. From Fig.~\hyperref[fig:results_TMSV]{\ref*{fig:results_TMSV}(b)}, we can conclude that $D_{\mathrm{I}}$ is the optimal NPT criterion for the TMSV state because it has the greatest statistical power for the full range of $\zeta$ considered. 
Notably, as $\zeta$ is increased beyond 1, the determinants $D_{\mathrm{I-III}}$ begin to perform more poorly and our results thus show that for stronger squeezing, it also becomes more challenging to confidently determine that the state is entangled. This behavior in statistical power will be discussed further below after we discuss finite-statistics error and open-system dynamics.

\begin{figure*}
\centering
\includegraphics[width=\textwidth]{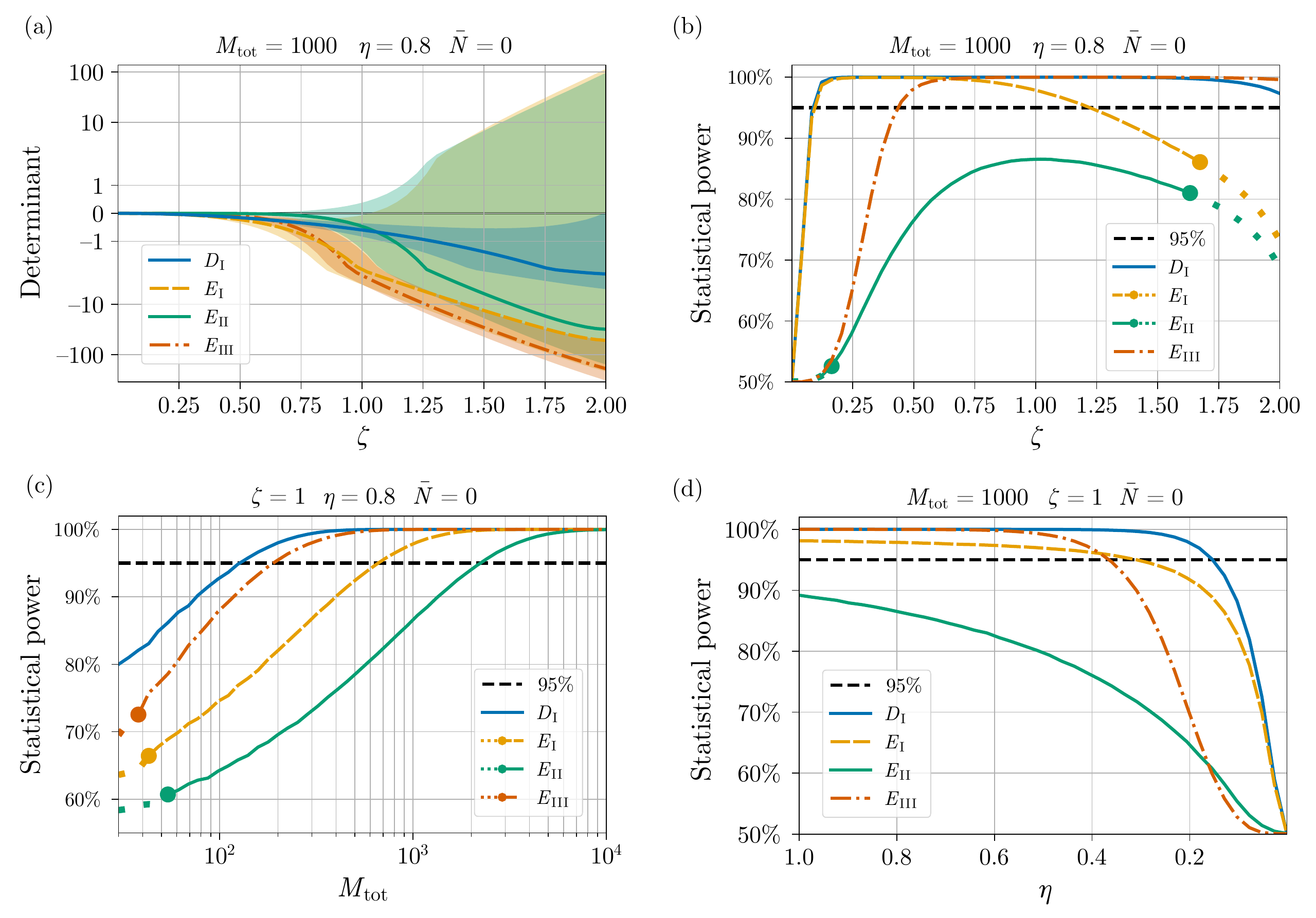}
\caption{NPT entanglement criteria for the single-quantum-subtracted TMSV state $\ket{\Psi}_{\mathrm{SUB}}$ explored using \textit{NPyT}. 
Here, we plot the NPT criteria---$D_{\mathrm{I}}$ and $E_{\mathrm{I}-\mathrm{III}}$---that can detect entanglement within the subset $d = 2$ and $n\leq 4$. We plot their statistical power as functions of the squeezing parameter $\zeta$, the loss parameter $\eta$, and the total number of measurements for each determinant $M_{\mathrm{tot}}$. The parameters used for each plot are given as plot headers.
(a) Determinants $D_{\mathrm{I}}$ and $E_{\mathrm{I}-\mathrm{III}}$ and their associated 2.5th to 97.5th percentile interval are plotted with the squeezing parameter over the range $0\leq\zeta\leq2$.
(b) Statistical power as $\zeta$ is varied for fixed $M_{\mathrm{tot}}$, $\eta$, and $\bar{N}$. We note that for small $\zeta$, $D_{\mathrm{I}}$ is just higher than $E_{\mathrm{I}}$. 
(c) Statistical power with $M_{\mathrm{tot}}$. Here, $\eta=0.8$, $\bar{N}=0$, and $\zeta=1$ have been chosen so $D_{\mathrm{I}}$ and $E_{\mathrm{I}-\mathrm{III}}$ exceed the 95\% threshold within the range $10^2\leq M_{\mathrm{tot}}\leq10^4$. For these parameters, $D_{\mathrm{I}}$ requires the fewest number of measurements to approach 100\% statistical power, while $E_{\mathrm{II}}$ requires the most. 
(d) Statistical power with loss parameter $\eta$ illustrating that $D_{\mathrm{I}}$ has the greatest robustness to loss.
}
\label{fig:tmsv_sub_results_fig}
\end{figure*}

To illustrate how the total number of measurements $M_{\mathrm{tot}}$ influences the entanglement test, 
in Fig.~\hyperref[fig:results_TMSV]{\ref*{fig:results_TMSV}(c)} we plot the statistical power
with $M_{\mathrm{tot}}$ for each $\textrm{det}[\mathbf{A}]$. 
Provided $\mathrm{det}[\mathbf{A}]<0$, it is evident that statistical power increases with $M_{\mathrm{tot}}$. 
Notably, we see that $D_{\mathrm{I}}$ demonstrates the best performance across the entire range of $M_{\mathrm{tot}}$ considered here. Also note here that the curves in this plot are not smooth due to the effects of integer steps in measurement number for any moment, see Section~\ref{sec:minimizing_error_on_NPT}.

Next, we examine the effects of interactions with the environment on the statistical power. In Fig.~\hyperref[fig:results_TMSV]{\ref*{fig:results_TMSV}(d)} we plot the statistical power with $\eta$. As $\eta$ decreases, the state loses more information to the environment and consequently the entanglement of the initial bipartite state degrades. For small values of $\eta$, the uncertainties on the determinants can be much larger than the magnitude of the determinants and hence the statistical power tends toward 50\%.
This is because as $\eta$ tends to 0, the values of $D_{\mathrm{I-III}}$ also tend towards 0, and the associated uncertainties are finite and span both the regions $\textrm{Re}\{\mathrm{det}[\bar{\mathbf{A}}]\}>0$ and $\textrm{Re}\{\mathrm{det}[\bar{\mathbf{A}}]\}<0$, leading to a 50\% statistical power.
We also qualitatively note that $D_{\mathrm{III}}$ performs the worst across the entire range of $\eta$ for the TMSV state. Notably, $D_{\mathrm{III}}$ is comprised of the most number of moments---see Table~\ref{tab:TMSV}. In contrast, $D_{\mathrm{I}}$, which is comprised of the fewest number of moments, shows the greatest robustness to optical loss. 

With both finite-statistics error and the effects of optical loss discussed, we now return to the observation in Fig.~\hyperref[fig:results_TMSV]{\ref*{fig:results_TMSV}(b)} that the statistical power is non-monotonic and reduces for large squeezing parameter. This may be counter-intuitive as one may expect that it is easier to confidently detect entanglement with increasing squeezing. However, our statistical framework reveals the opposite and it is in fact more challenging to confidently detect entanglement for large squeezing due to a combination of finite-statistics error and open-system dynamics. Indeed, the trend observed in Fig.~\hyperref[fig:results_TMSV]{\ref*{fig:results_TMSV}(b)} where the statistical power initially increases with $\zeta$ and then decreases for large $\zeta$ still remains when either $M_{\mathrm{tot}}$ or $\eta$ is increased. And, it is important to note that this behavior remains as $\eta \rightarrow 1$ for a finite $M_{\mathrm{tot}}$ demonstrating that reduced statistical power is not purely due to vacuum noise and the important role of total measurement number. Thus, for increasing squeezing parameter, more measurements are needed to confidently detect the entanglement, which may be intuitively understood as correlations now exist over a greater region of phase space to be sampled.

Comparing the performance of $D_{\mathrm{I-III}}$, we find that choosing the NPT criterion with the fewest number of moments gives the greatest entanglement-detection confidence. Notably, other NPT criteria with $d=2$, other than $D_{\mathrm{I}}$, are not capable of detecting entanglement for the TMSV state. To see this, consider $\eta=1$, in which case $\expval{\hat{a}^\dag \hat{a}}=\expval{\hat{b}^\dag \hat{b}}=\sinh{(\zeta/2)}^2$, $\expval{\hat{a}\hat{b}}=\expval{\hat{a}^\dag \hat{b}^\dag}=-\cosh{(\zeta/2)}\sinh{(\zeta/2)}$, and all other unique second-order moments are zero. By calculating determinants of order $n=2$, as described in Appendix~\ref{app:entanglement_criteria}, one may see that, in the absence of loss, $D_{\mathrm{I}}=\expval{\hat{a}^\dag \hat{a}}\expval{\hat{b}^\dag \hat{b}}-\expval{\hat{a}\hat{b}}\expval{\hat{a}^\dag \hat{b}^\dag}=-\sinh{(\zeta/2)}^2$ is the only negative determinant constructed from a $2\times2$ submatrix. As the inclusion of optical loss reduces the negativity of the determinant, $D_{\mathrm{I}}$ is the only determinant parametrized by $n=2$ and $d=2$ that is capable of detecting the entanglement of the TMSV state. Finally, we would like to highlight that among all the determinants of order $n=2$, which includes Simon's criterion $D_{\mathrm{III}}$, determinant $D_{\mathrm{I}}$ shows the best performance, which we understand to be due to its small dimension and off-diagonal matrix elements that access the correlation terms.

\subsubsection{Quanta-subtracted two-mode squeezed vacuum state}\label{sec:subTMSV}

\begin{figure*}
\centering
\includegraphics[width=\textwidth]{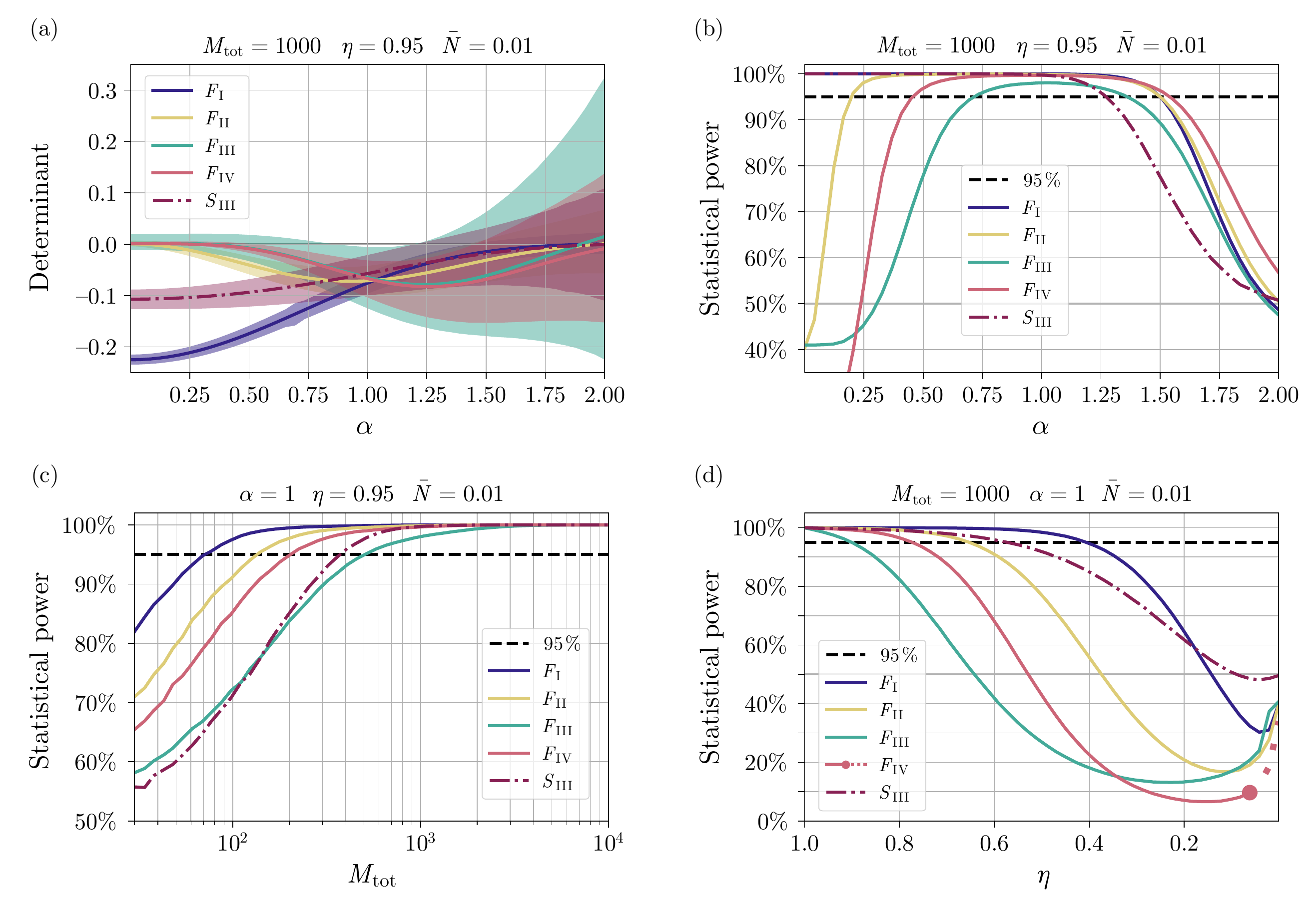}
\caption{
NPT entanglement criteria for the two-mode Schr{\"o}dinger-cat state $\ket{\Psi}_{\mathrm{CAT}}$ explored using \textit{NPyT}. 
Here, we plot the NPT criteria
$F_{\mathrm{I}-\mathrm{IV}}$, the determinants $\mathrm{det}[\mathbf{A}]$ that are capable of detecting entanglement within the $d=2$ and $n\leq 4$ subset. For comparison, we also plot $S_{\mathrm{III}}$ defined in Eq.~\eqref{eq:S3_matrix}, which has $d=3$ and $n=4$.
We plot these NPT criteria and their statistical power with cat-state amplitude $\alpha$, the total number of measurements allocated to each determinant $M_{\mathrm{tot}}$, and the beam-splitter parameter $\eta$ that quantifies environmental interactions. Here, we use a non-zero thermal occupation of the environment, relevant for e.g. mechanical systems, and the parameters used for each plot are given as plot headers.
(a)~The determinants 
and their associated 2.5th to 97.5th percentile interval, plotted with cat-state amplitude over the range $0\leq\alpha\leq2$.
(b) Statistical power with $\alpha$. For the parameters used, $F_{\mathrm{I}}$ and $S_{\mathrm{III}}$ are optimal for small $\alpha$ and $F_{\mathrm{IV}}$ performs better for large $\alpha$.
(c) Statistical power with $M_{\mathrm{tot}}$.
For the parameters used here, $F_{\mathrm{I}-\mathrm{IV}}$ and $S_{\mathrm{III}}$ all exceed 95\% statistical power within the range $10^2\leq M_{\mathrm{tot}}\leq10^4$. Here, $F_{\mathrm{I}}$ requires the fewest number of measurements to approach 100\% statistical power, while $F_{\mathrm{III}}$ requires the most. 
(d) Statistical power with $\eta$. For a lossless system with $\eta=1$, all determinants approach 100\% statistical power but as losses increase all the determinants degrade. Notably, $F_{\mathrm{I}}$ is capable of withstanding significant coupling to the thermal environment and maintains $>95\%$ statistical power for $\eta \gtrsim 0.4$.}
\label{fig:results_TMSC}
\end{figure*}

Here, we conduct the same analysis performed in the previous section but for the state $\ket{\Psi}_{\mathrm{SUB}}$, which is generated by subtracting one quantum of energy from each mode of a TMSV state. Mathematically, this state is given by setting $l=1$ and $k=1$ in Eq.~\eqref{eq:subTMSV}. 
In
Fig.~\hyperref[fig:tmsv_sub_results_fig]{\ref*{fig:tmsv_sub_results_fig}(a)} we plot the determinants that exhibit negativity as we vary the squeezing parameter $\zeta$. These successful determinants include $D_{\mathrm{I}}$ from Section~\ref{sec:TMSV} and also the determinants labeled by $E_{\mathrm{I}-\mathrm{V}}$, which are summarized in Table~\ref{tab:TMSV_sub}. These determinants are found by first calculating all $105$ possible matrices $\mathbf{A}$ which satisfy $d=2$ and $n\leq 4$~\footnote{{S}ubmatrices with $n\leq 4$ are generated by keeping rows and columns of $\mathbf{M}$ up to $i,j\leq15$. Thus, there are $\bigl(\begin{smallmatrix}15\\2\end{smallmatrix}\bigr)=105$ matrices with $d=2$ and $n\leq 4$.}, and then identifying the subset for which $\mathrm{det}[\mathbf{A}]<0$ for any value of the squeezing parameter in the range $0\leq\zeta\leq2$. 
By swapping $\hat{a}$ ($\hat{a}^\dag$) with $\hat{b}$ ($\hat{b}^\dag$), we find that $E_{\mathrm{I}}=E_{\mathrm{IV}}$ and $E_{\mathrm{II}}=E_{\mathrm{V}}$ and therefore we plot the only unique determinants $D_{\mathrm{I}}$ and $E_{\mathrm{I}-\mathrm{III}}$ capable of detecting entanglement in Fig.~\ref{fig:tmsv_sub_results_fig}. 
As in Section~\ref{sec:TMSV}, we choose $\eta=0.8$, and we again assume the environment is well described by the vacuum state for optical fields, thus $\bar{N}=0$. However, in contrast to Section~\ref{sec:TMSV}, we choose $M_{\mathrm{tot}}=1000$ to reduce the comparatively large uncertainties in the determinants $E_{\mathrm{I}}$ and $E_\mathrm{II}$, which include fourth-order moments. It is also curious to note here that determinants $D_{\mathrm{II}}$ and $D_{\mathrm{III}}$ studied in the previous section to detect entanglement for the TMSV state are not able to detect entanglement for the single-quantum-subtracted TMSV state.

In Fig.~\hyperref[fig:tmsv_sub_results_fig]{\ref*{fig:tmsv_sub_results_fig}(b)}, we plot the entanglement test statistical power 
with $\zeta$. For $\zeta\lesssim0.75$ we observe $D_{\mathrm{I}}$ and $E_{\mathrm{I}}$ having the highest statistical powers, and for $\zeta\gtrsim0.75$ we find that $D_{\mathrm{I}}$ and $E_{\mathrm{III}}$ can detect entanglement with the greatest statistical power. In Fig.~\hyperref[fig:tmsv_sub_results_fig]{\ref*{fig:tmsv_sub_results_fig}(c)}, we plot the statistical power as a function of the total number of measurements $M_{\mathrm{tot}}$. Compared to Fig.~\hyperref[fig:results_TMSV]{\ref*{fig:results_TMSV}(c)}, we consider a larger range of $M_{\mathrm{tot}}$ in order to observe $E_{\mathrm{II}}$ exceed a statistical power of $95\%$. Here, we find that as $M_{\mathrm{tot}}$ is increased, $D_{\mathrm{I}}$ is the first to reach a statistical power of 95\%. In Fig.~\hyperref[fig:tmsv_sub_results_fig]{\ref*{fig:tmsv_sub_results_fig}(d)}, we plot statistical power with $\eta$. As $\eta$ is reduced from 1 to 0 the statistical power of $D_{\mathrm{I}}$ and $E_{\mathrm{I}-\mathrm{III}}$ tends towards 50\%, as in Section~\ref{sec:TMSV}. Notably, as in Section~\ref{sec:TMSV}, $D_{\mathrm{I}}$ is the most robust NPT criterion to environmental interactions. 
Thus, we conclude that $D_{\mathrm{I}}$ shows high statistical power across the parameter range investigated for the single-quantum-subtracted TMSV state with $d=2$ and $n\leq 4$, and we note that $E_{\mathrm{III}}$, which has a larger $n$, becomes optimal for high squeezing.

\subsubsection{Two-mode Schr{\"o}dinger-cat state}
\label{sec:TMSC}

As a final example, we use our framework to find the optimal NPT test for the two-mode Schr{\"o}dinger-cat state $\ket{\Psi}_{\mathrm{CAT}}\propto \ket{\alpha}\ket{0}-\ket{0}\ket{\alpha}$ under the constraint $d=2$ and $n\leq 4$. For these constraints, \textit{NPyT} identifies six successful determinants that we label $F_{\mathrm{I}-\mathrm{VI}}$, summarized in Table~\ref{tab:TMSC}. 
Owing to the symmetry between the two subsystems of this state, swapping the field operators $\hat{a}$ ($\hat{a}^{\dag}$) and $\hat{b}$ ($\hat{b}^{\dag}$) gives $F_{\mathrm{II}}=F_{\mathrm{V}}$ and $F_{\mathrm{III}}=F_{\mathrm{VI}}$, leaving four unique determinants for these constraints.

In Fig.~\hyperref[fig:results_TMSC]{\ref*{fig:results_TMSC}(a)} we plot the NPT entanglement tests identified along with test $S_{\text{III}}$ defined in Eq.~\eqref{eq:S3_matrix}. Here, the beam-splitter parameter is $\eta=0.95$, the thermal occupation number of the bath is $\bar{N}=0.01$, and the total number of measurements is $M_{\mathrm{tot}}=1000$. 
Although the determinant $S_{\mathrm{III}}$ is more experimentally demanding to measure due to its higher dimension $d=3$, we compare it here as it has been previously highlighted in the context of detecting entanglement for two-mode cat states, see Section~\ref{sec:characterizing} and our previous work in Ref.~\cite{Kanari2022}. We also note that $F_{\mathrm{I}}<0$ falls within the class of Hillery--Zubairy entanglement criteria~\cite{hillery2006entanglement, HilleryZubairy2006PRA}. 

To further investigate these NPT criteria, in Fig.~\hyperref[fig:results_TMSC]{\ref*{fig:results_TMSC}(b)} we plot the statistical power with of $\alpha$. For the parameters considered, we can conclude that $F_{\mathrm{I}}$ and $S_{\mathrm{III}}$ are the optimal NPT criteria for states with smaller $\alpha$, whereas, $F_{\mathrm{IV}}$ has the highest statistical power for the largest values of $\alpha$ explored. 
These trends, and comparing $F_{\mathrm{I}}$ with $F_{\mathrm{IV}}$ and $S_{\mathrm{III}}$, suggest that low $n$ and low $d$ suffice for low $\alpha$ and it may be worthwhile exploring larger $n$ and $d$ for higher $\alpha$.

In Fig.~\hyperref[fig:results_TMSC]{\ref*{fig:results_TMSC}(c)} we plot statistical power with the total measurement number $M_{\mathrm{tot}}$ allocated to each determinant. 
Notably, here we see that $F_{\mathrm{I}}$ is first to reach a statistical power of 95\% with increasing $M_{\mathrm{tot}}$. 
In Fig.~\hyperref[fig:results_TMSC]{\ref*{fig:results_TMSC}(d)} we plot the statistical power with $\eta$. 
We observe that $F_{\mathrm{III}}$ and $F_{\mathrm{IV}}$ degrade the most quickly as $\eta$ is decreased and we note these are the only determinants where all four entries are comprised of fourth-order moments, helping to demonstrate that determinants composed of increasingly high $n$ are more susceptible to open-system dynamics. By contrast, $F_{\mathrm{I}}$ shows the greatest robustness to decreasing $\eta$ and is comprised of lower order moments---see Table~\ref{tab:TMSC}. For the parameters considered here, we therefore identify $F_{\mathrm{I}}$ as the optimal NPT criterion for a two-mode Schr\"odinger-cat state owing to its high statistical power, favorable scaling with measurement number, and resilience to open-system dynamics. Looking beyond to larger values of the cat-state amplitude $\alpha$, the test $F_{\mathrm{IV}}$ becomes the optimal choice, and tests with larger $n$ or $d$ may also provide better performance in this regime.

\section{Conclusions}
We have developed a statistical framework for selecting the optimal bipartite continuous-variable entanglement test via the negative partial transpose. Experimental uncertainties are accounted for by considering open-system dynamics and a fixed resource in the total number of measurements. Using a Lagrange-multiplier method, our framework enables the uncertainty on a general NPT criterion to be minimized via an optimally weighted allocation of measurement resources to the observables.
Maximizing the statistical power of a hypothesis test over a set of NPT criteria then allows the optimal criterion to be selected for a given state of interest. 
To demonstrate the utility of our scheme, we applied our framework to three example states including the two-mode squeezed vacuum state, the single-quantum-subtracted two-mode squeezed vacuum state, and the two-mode Schr{\"o}dinger-cat state. For each of these states, we identified the optimal NPT criterion in the presence of open-system dynamics as the total number of measurements and system parameters are varied.

An important trend we observe is that the determinant distributions become more heavy-tailed as the dimension of the test $d$ increases causing the statistical power to decrease. This behavior is due to the increase in the nonlinearity of the determinant function with $d$, i.e. a product of $d$ random variables is required to obtain the determinant and hence the determinant becomes more non-Gaussian with increasing $d$.

Our method is applicable to a wide range of bipartite continuous-variable systems including optical~\cite{ou1992realization,furusawa1998unconditional,bowen2003experimental,eisenberg2004quantum,afek2010high} and microwave fields~\cite{rauschenbeutel2001controlled, wang2011deterministic,Wang2016}, massive mechanical systems~\cite{Mancini2002,lee2011entangling,clarke2020generating,neveu2021preparation,mercier2021quantum,kotler2021direct}, motional states of trapped ions~\cite{jost2009entangled,zhang2018noon,jeon2024experimental}, and entanglement between different physical systems~\cite{palomaki2013entangling, xia2014opto, thomas2021entanglement}.
Indeed, as experiments advance towards generating highly non-Gaussian and macroscopic entangled states, the verification of entanglement will become more challenging due to the increasingly detrimental effect of environmental interactions. 
For example, in quantum optomechanics, such experiments include studies of multimode states in the quantum regime~\cite{nielsen2017multimode}, the nonlinear regime~\cite{vanner2011selective, brawley2016nonlinear, leijssen2017nonlinear, clarke2023cavity}, explorations of gravity-induced entanglement~\cite{qvarfort2020mesoscopic,miki2022non,plato2023enhanced,bose2023massive}, and tests of unconventional decoherence mechanisms with entangled systems~\cite{furry1936note, kiesewetter2017pulsed}.
Our framework may be used to advance such experimental efforts by helping to select the optimal NPT test to experimentally perform for a known state. Alternatively, for an unknown or partially known state, our framework may be adapted so that empirical data are used, instead of simulated moments of a known state, to select the optimal test in post-processing or to inform which subset of moments to focus on for a subsequent more careful experiment. For instance, assuming an initially unknown Gaussian state, and then all the quadratic moments are measured, there are still 31 different NPT criteria to choose between and our framework can select the test with the least statistical error to use.

Beyond applicability to NPT-based bipartite entanglement tests, our framework may be extended to identify optimal tests of non-classicality in single-mode quantum states~\cite{ShchukinNonClass}, multi-mode NPT entanglement~\cite{shchukin2006conditions, Miranowicz2009, miranowicz2010testing}, and may help identify NPT bound entanglement~\cite{horodecki1998mixed}.

\vspace{1mm}
\begin{acknowledgments}
\textit{Acknowledgements.}---We acknowledge useful discussions with R. Clarke, E.~A. Cryer-Jenkins, and C. Michaelidou. 
This project was supported by UK Research and Innovation (Grants No. MR/S032924/1 and No. MR/X024105/1), the Engineering and Physical Sciences Research Council (Grant No. EP/L016524/1), and the Science and Technology Facilities Council (Grant No. ST/W006553/1).
SQ is funded by the Wallenberg Initiative on Networks and Quantum Information (WINQ) and by the Marie Skłodowska–Curie Action IF Programme (Grant No. 101027183). Nordita is supported by NordForsk. 
\end{acknowledgments}

\section*{Data availability statement}

To implement our framework, we have developed the \textit{\textbf{N}PT test optimization \textbf{Py}thon \textbf{T}oolbox} (\textit{NPyT}). This toolbox, along with the code for the three example states, can be found in the Quantum Measurement Lab GitHub repository~\cite{lydia_code}. 
\textit{NPyT} uses functions from the {QuTiP} package~\cite{johansson2012qutip}.

\bibliography{bibliography_allblack}

@article{Adesso2007,
  title={Entanglement in continuous-variable systems: recent advances and current perspectives},
  author={Adesso, Gerardo and Illuminati, Fabrizio},
  journal={Journal of Physics A: Mathematical and Theoretical},
  volume={40},
  number={28},
  pages={7821},
  year={2007},
  publisher={IOP Publishing},
  url={https://doi.org/10.1088/1751-8113/40/28/S01}
}

@article{birrittella2023unconditional,
  title = {Unconditional remote entanglement using second-harmonic generation and twin two-mode squeezed vacuum states},
  author = {Birrittella, Richard J. and Schneeloch, James and Tison, Christopher C. and Fanto, Michael L. and Alsing, Paul M. and Gerry, Christopher C.},
  journal = {Physical Review A},
  volume = {107},
  issue = {3},
  pages = {033505},
  numpages = {11},
  year = {2023},
  month = {Mar},
  publisher = {American Physical Society},
  url = {https://link.aps.org/doi/10.1103/PhysRevA.107.033505}
}

@article{horodecki1996necessary,
  title={On the necessary and sufficient conditions for separability of mixed quantum states},
  author={Horodecki, Michal and Horodecki, Pawel and Horodecki, Ryszard},
  journal={Physics Letters A},
  volume={223},
  number={1},
  year={1996},
  publisher={Citeseer},
  url={https://citeseerx.ist.psu.edu/document?repid=rep1&type=pdf&doi=f829c395a281fde3395edaff91469190aa77dbc3}
}

@article{horodecki2009quantum,
  title={Quantum entanglement},
  author={Horodecki, Ryszard and Horodecki, Pawe{\l} and Horodecki, Micha{\l} and Horodecki, Karol},
  journal={Reviews of Modern Physics},
  volume={81},
  number={2},
  pages={865},
  year={2009},
  publisher={APS},
  url={https://doi.org/10.1103/RevModPhys.81.865}
}

@article{Miranowicz2009,
  title={Inseparability criteria based on matrices of moments},
  author={Miranowicz, Adam and Piani, Marco and Horodecki, Pawe{\l} and Horodecki, Ryszard},
  journal={Physical Review A},
  volume={80},
  number={5},
  pages={052303},
  year={2009},
  publisher={APS},
  url={https://doi.org/10.1103/PhysRevA.80.052303}
}

@article{Zhang2021,
  title={Hierarchy of Nonlinear Entanglement Dynamics for Continuous Variables},
  author={Zhang, Da and Barral, David and Cai, Yin and Zhang, Yanpeng and Xiao, Min and Bencheikh, Kamel},
  journal={Physical Review Letters},
  volume={127},
  number={15},
  pages={150502},
  year={2021},
  publisher={APS},
  url={https://doi.org/10.1103/PhysRevLett.127.150502}
}

@article{Peres1996,
  title = {Separability Criterion for Density Matrices},
  author = {Peres, Asher},
  journal = {Physical Review Letters},
  volume = {77},
  issue = {8},
  pages = {1413--1415},
  numpages = {0},
  year = {1996},
  month = {Aug},
  publisher = {American Physical Society},
  doi = {10.1103/PhysRevLett.77.1413}
}

@article{Horodecki1996,
title = "Separability of mixed states: necessary and sufficient conditions",
journal = "Physics Letters A",
volume = "223",
number = "1",
pages = "1 - 8",
year = "1996",
doi = "https://doi.org/10.1016/S0375-9601(96)00706-2",
author = "Michał Horodecki and Paweł Horodecki and Ryszard Horodecki"
}

@article{Simon2000,
  title = {Peres-Horodecki Separability Criterion for Continuous Variable Systems},
  author = {Simon, R.},
  journal = {Physical Review Letters},
  volume = {84},
  issue = {12},
  pages = {2726--2729},
  numpages = {0},
  year = {2000},
  month = {Mar},
  publisher = {American Physical Society},
  doi = {10.1103/PhysRevLett.84.2726},
  url = {https://link.aps.org/doi/10.1103/PhysRevLett.84.2726}
}

@article{Shchukin2005,
  title = {Inseparability Criteria for Continuous Bipartite Quantum States},
  author = {Shchukin, E. and Vogel, W.},
  journal = {Physical Review Letters},
  volume = {95},
  issue = {23},
  pages = {230502},
  numpages = {4},
  year = {2005},
  month = {Nov},
  publisher = {American Physical Society},
  doi = {10.1103/PhysRevLett.95.230502}
}

@article{milburn2016nonclassical,
  title = {Nonclassical-state generation in macroscopic systems via hybrid discrete-continuous quantum measurements},
  author = {Milburn, T. J. and Kim, M. S. and Vanner, M. R.},
  journal = {Phys. Rev. A},
  volume = {93},
  issue = {5},
  pages = {053818},
  numpages = {11},
  year = {2016},
  month = {May},
  publisher = {American Physical Society},
  doi = {10.1103/PhysRevA.93.053818},
  url = {https://link.aps.org/doi/10.1103/PhysRevA.93.053818}
}

@article{Kanari2022,
  title={Two-mode Schr{\"o}dinger-cat states with nonlinear optomechanics: generation and verification of non-Gaussian mechanical entanglement},
  author={Kanari-Naish, Lydia A and Clarke, Jack and Qvarfort, Sofia and Vanner, Michael R},
  journal={Quantum Science and Technology},
  volume={7},
  number={3},
  pages={035012},
  year={2022},
  publisher={IOP Publishing},
  url={https://doi.org/10.1088/2058-9565/ac6dfd}
}

@article{Agarwal2005,
  title={Inseparability inequalities for higher order moments for bipartite systems},
  author={Agarwal, Girish Saran and Biswas, Asoka},
  journal={New Journal of Physics},
  volume={7},
  number={1},
  pages={211},
  year={2005},
  publisher={IOP Publishing},
  url={https://doi.org/10.1088/1367-2630/7/1/211}
}

@article{hillery2006entanglement,
  title={Entanglement conditions for two-mode states},
  author={Hillery, Mark and Zubairy, M Suhail},
  journal={Physical Review Letters},
  volume={96},
  number={5},
  pages={050503},
  year={2006},
  publisher={APS},
  url={https://journals.aps.org/prl/abstract/10.1103/PhysRevLett.80.5239}
}

@article{HilleryZubairy2006PRA,
  title = {Entanglement conditions for two-mode states: Applications},
  author = {Hillery, Mark and Zubairy, M. Suhail},
  journal = {Physical Review A},
  volume = {74},
  issue = {3},
  pages = {032333},
  numpages = {7},
  year = {2006},
  month = {Sep},
  publisher = {American Physical Society},
  doi = {10.1103/PhysRevA.74.032333},
  url = {https://link.aps.org/doi/10.1103/PhysRevA.74.032333}
}

@article{Joo2011,
  title = {Quantum Metrology with Entangled Coherent States},
  author = {Joo, Jaewoo and Munro, William J. and Spiller, Timothy P.},
  journal = {Physical Review Letters},
  volume = {107},
  issue = {8},
  pages = {083601},
  numpages = {4},
  year = {2011},
  month = {Aug},
  publisher = {American Physical Society},
  doi = {10.1103/PhysRevLett.107.083601}
}

@article{Wang2016,
	author = {Wang, Chen and others},
	title = {A {S}chr{\"o}dinger cat living in two boxes},
	volume = {352},
	number = {6289},
	pages = {1087--1091},
	year = {2016},
	doi = {10.1126/science.aaf2941},
	publisher = {American Association for the Advancement of Science},
	journal = {Science}
}

@article{Yadin2016,
  title = {General framework for quantum macroscopicity in terms of coherence},
  author = {Yadin, Benjamin and Vedral, Vlatko},
  journal = {Physical Review A},
  volume = {93},
  issue = {2},
  pages = {022122},
  numpages = {7},
  year = {2016},
  month = {Feb},
  publisher = {American Physical Society},
  doi = {10.1103/PhysRevA.93.022122}
}

@article{Marshall2003,
  title={Towards quantum superpositions of a mirror},
  author={Marshall, William and Simon, Christoph and Penrose, Roger and Bouwmeester, Dik},
  journal={Physical Review Letters},
  volume={91},
  number={13},
  pages={130401},
  year={2003},
  publisher={APS},
  url={https://doi.org/10.1103/PhysRevLett.91.130401}
}

@article{Bose1999,
  title = {Scheme to probe the decoherence of a macroscopic object},
  author = {Bose, S. and Jacobs, K. and Knight, P. L.},
  journal = {Physical Review A},
  volume = {59},
  issue = {5},
  pages = {3204--3210},
  numpages = {0},
  year = {1999},
  month = {May},
  publisher = {American Physical Society},
  doi = {10.1103/PhysRevA.59.3204}
}

@article{Kleckner2008,
  title={Creating and verifying a quantum superposition in a micro-optomechanical system},
  author={Kleckner, Dustin and Pikovski, Igor and Jeffrey, Evan and Ament, Luuk and Eliel, Eric and Van Den Brink, Jeroen and Bouwmeester, Dirk},
  journal={New Journal of Physics},
  volume={10},
  number={9},
  pages={095020},
  year={2008},
  publisher={IOP Publishing},
  doi={10.1088/1367-2630/10/9/095020}
}

@article{LeeCW2011,
  title = {Quantification of Macroscopic Quantum Superpositions within Phase Space},
  author = {Lee, Chang-Woo and Jeong, Hyunseok},
  journal = {Physical Review Letters},
  volume = {106},
  issue = {22},
  pages = {220401},
  numpages = {4},
  year = {2011},
  month = {May},
  publisher = {American Physical Society},
  doi = {10.1103/PhysRevLett.106.220401}
}

@article{Duan2000,
  title = {Inseparability Criterion for Continuous Variable Systems},
  author = {Duan, Lu-Ming and Giedke, G. and Cirac, J. I. and Zoller, P.},
  journal = {Physical Review Letters},
  volume = {84},
  issue = {12},
  pages = {2722--2725},
  numpages = {0},
  year = {2000},
  month = {Mar},
  publisher = {American Physical Society},
  doi = {10.1103/PhysRevLett.84.2722}
}

@article{Mancini2002,
  title = {Entangling Macroscopic Oscillators Exploiting Radiation Pressure},
  author = {Mancini, Stefano and Giovannetti, Vittorio and Vitali, David and Tombesi, Paolo},
  journal = {Physical Review Letters},
  volume = {88},
  issue = {12},
  pages = {120401},
  numpages = {4},
  year = {2002},
  month = {Mar},
  publisher = {American Physical Society},
  doi = {10.1103/PhysRevLett.88.120401}
}

@article{Raymer2003,
  title = {Separability criterion for separate quantum systems},
  author = {Raymer, M. G. and Funk, A. C. and Sanders, B. C. and de Guise, H.},
  journal = {Physical Review A},
  volume = {67},
  issue = {5},
  pages = {052104},
  numpages = {5},
  year = {2003},
  month = {May},
  publisher = {American Physical Society},
  doi = {10.1103/PhysRevA.67.052104}
}

@article{ShchukinNonClass,
  title = {Nonclassicality criteria in terms of moments},
  author = {Shchukin, E. and Richter, Th. and Vogel, W.},
  journal = {Physical Review A},
  volume = {71},
  issue = {1},
  pages = {011802},
  numpages = {4},
  year = {2005},
  month = {Jan},
  publisher = {American Physical Society},
  doi = {10.1103/PhysRevA.71.011802},
  url = {https://link.aps.org/doi/10.1103/PhysRevA.71.011802}
}

@article{Navarrete2012,
  title = {Enhancing quantum entanglement by photon addition and subtraction},
  author = {Navarrete-Benlloch, Carlos and Garc\'{\i}a-Patr\'on, Ra\'ul and Shapiro, Jeffrey H. and Cerf, Nicolas J.},
  journal = {Physical Review A},
  volume = {86},
  issue = {1},
  pages = {012328},
  numpages = {9},
  year = {2012},
  month = {Jul},
  publisher = {American Physical Society},
  doi = {10.1103/PhysRevA.86.012328},
  url = {https://link.aps.org/doi/10.1103/PhysRevA.86.012328}
}

@book{Leonhardt1997, 
    place={Cambridge}, 
    title={6}, 
    booktitle={Measuring the Quantum State of Light}, 
    publisher={Cambridge University Press}, 
    author={Leonhardt, Ulf}, 
    year={1997},
    url={https://www.google.co.uk/books/edition/Measuring_the_Quantum_State_of_Light/wmsJy1A_cyIC?hl=en&gbpv=0}
}

@article{Dell2006,
  title={Test of inseparability criteria for squeezed number states of the radiation field},
  author={Dell’Anno, Fabio and Siena, Silvio De and Illuminati, Fabrizio},
  journal={Open Syst. Inf. Dyn.},
  volume={13},
  number={4},
  pages={383--392},
  year={2006},
  publisher={Springer},
  url={https://doi.org/10.1007/s11080-006-9020-4}
}

@article{Barnett2018,
  title = {Statistics of photon-subtracted and photon-added states},
  author = {Barnett, Stephen M. and Ferenczi, Gergely and Gilson, Claire R. and Speirits, Fiona C.},
  journal = {Physical Review A},
  volume = {98},
  issue = {1},
  pages = {013809},
  numpages = {13},
  year = {2018},
  month = {Jul},
  publisher = {American Physical Society},
  doi = {10.1103/PhysRevA.98.013809},
  url = {https://link.aps.org/doi/10.1103/PhysRevA.98.013809}
}

@article{Vanner2013,
  title = {Quantum State Orthogonalization and a Toolset for Quantum Optomechanical Phonon Control},
  author = {Vanner, M. R. and Aspelmeyer, M. and Kim, M. S.},
  journal = {Physical Review Letters},
  volume = {110},
  issue = {1},
  pages = {010504},
  numpages = {5},
  year = {2013},
  month = {Jan},
  publisher = {American Physical Society},
  doi = {10.1103/PhysRevLett.110.010504},
  url = {https://link.aps.org/doi/10.1103/PhysRevLett.110.010504}
}

@article{Enzian2021,
  title = {Single-Phonon Addition and Subtraction to a Mechanical Thermal State},
  author = {Enzian, G. and Price, J. J. and Freisem, L. and Nunn, J. and Janousek, J. and Buchler, B. C. and Lam, P. K. and Vanner, M. R.},
  journal = {Physical Review Letters},
  volume = {126},
  issue = {3},
  pages = {033601},
  numpages = {6},
  year = {2021},
  month = {Jan},
  publisher = {American Physical Society},
  doi = {10.1103/PhysRevLett.126.033601},
  url = {https://link.aps.org/doi/10.1103/PhysRevLett.126.033601}
}

@article{Enzian2021nonG,
  title = {Non-Gaussian Mechanical Motion via Single and Multiphonon Subtraction from a Thermal State},
  author = {Enzian, G. and Freisem, L. and Price, J. J. and Svela, A. \O{}. and Clarke, J. and Shajilal, B. and Janousek, J. and Buchler, B. C. and Lam, P. K. and Vanner, M. R.},
  journal = {Physical Review Letters},
  volume = {127},
  issue = {24},
  pages = {243601},
  numpages = {6},
  year = {2021},
  month = {Dec},
  publisher = {American Physical Society},
  doi = {10.1103/PhysRevLett.127.243601},
  url = {https://link.aps.org/doi/10.1103/PhysRevLett.127.243601}
}

@article{Ourjoumtsev2006,
  title={Generating optical Schrodinger kittens for quantum information processing},
  author={Ourjoumtsev, Alexei and Tualle-Brouri, Rosa and Laurat, Julien and Grangier, Philippe},
  journal={Science},
  volume={312},
  number={5770},
  pages={83--86},
  year={2006},
  publisher={American Association for the Advancement of Science},
  url={https://doi.org/10.1126/science.1122858}
}

@article{Neergaard2006,
  title={Generation of a superposition of odd photon number states for quantum information networks},
  author={Neergaard-Nielsen, Jonas S and Nielsen, B Melholt and Hettich, C and M{\o}lmer, Klaus and Polzik, Eugene S},
  journal={Physical Review Letters},
  volume={97},
  number={8},
  pages={083604},
  year={2006},
  publisher={APS},
  url={https://doi.org/10.1103/PhysRevLett.97.083604}
}

@article{Vedral1997,
  title = {Quantifying Entanglement},
  author = {Vedral, V. and Plenio, M. B. and Rippin, M. A. and Knight, P. L.},
  journal = {Physical Review Letters},
  volume = {78},
  issue = {12},
  pages = {2275--2279},
  numpages = {0},
  year = {1997},
  month = {Mar},
  publisher = {American Physical Society},
  doi = {10.1103/PhysRevLett.78.2275},
  url = {https://link.aps.org/doi/10.1103/PhysRevLett.78.2275}
}

@article{shchukin2006conditions,
  title={Conditions for multipartite continuous-variable entanglement},
  author={Shchukin, Evegny and Vogel, Werner},
  journal={Physical Review A},
  volume={74},
  number={3},
  pages={030302},
  year={2006},
  publisher={APS},
  url={https://doi.org/10.1103/PhysRevA.74.030302}
}

@article{ockeloen2018stabilized,
  title={Stabilized entanglement of massive mechanical oscillators},
  author={Ockeloen-Korppi, CF and Damsk{\"a}gg, E and Pirkkalainen, J-M and Asjad, M and Clerk, AA and Massel, F and Woolley, MJ and Sillanp{\"a}{\"a}, MA},
  journal={Nature},
  volume={556},
  number={7702},
  pages={478--482},
  year={2018},
  publisher={Nature Publishing Group UK London},
  url={https://doi.org/10.1038/s41586-018-0038-x}
}

@article{mercier2021quantum,
  title={Quantum mechanics--free subsystem with mechanical oscillators},
  author={Mercier de L{\'e}pinay, Laure and Ockeloen-Korppi, Caspar F and Woolley, Matthew J and Sillanp{\"a}{\"a}, Mika A},
  journal={Science},
  volume={372},
  number={6542},
  pages={625--629},
  year={2021},
  publisher={American Association for the Advancement of Science},
  url={https://doi.org/10.1126/science.abf5389}
}

@article{kotler2021direct,
  title={Direct observation of deterministic macroscopic entanglement},
  author={Kotler, Shlomi and Peterson, Gabriel A and Shojaee, Ezad and Lecocq, Florent and Cicak, Katarina and Kwiatkowski, Alex and Geller, Shawn and Glancy, Scott and Knill, Emanuel and Simmonds, Raymond W and others},
  journal={Science},
  volume={372},
  number={6542},
  pages={622--625},
  year={2021},
  publisher={American Association for the Advancement of Science},
  url={https://doi.org/10.1126/science.abf2998}
}

@article{lee2011entangling,
  title={Entangling macroscopic diamonds at room temperature},
  author={Lee, Ka Chung and Sprague, Michael R and Sussman, Benjamin J and Nunn, Joshua and Langford, Nathan K and Jin, X-M and Champion, Tessa and Michelberger, Patrick and Reim, Klaus F and England, Duncan and others},
  journal={Science},
  volume={334},
  number={6060},
  pages={1253--1256},
  year={2011},
  publisher={American Association for the Advancement of Science},
  url={https://doi.org/10.1126/science.1211914}
}

@article{thomas2021entanglement,
  title={Entanglement between distant macroscopic mechanical and spin systems},
  author={Thomas, Rodrigo A and Parniak, Micha{\l} and {\O}stfeldt, Christoffer and M{\o}ller, Christoffer B and B{\ae}rentsen, Christian and Tsaturyan, Yeghishe and Schliesser, Albert and Appel, J{\"u}rgen and Zeuthen, Emil and Polzik, Eugene S},
  journal={Nature Physics},
  volume={17},
  number={2},
  pages={228--233},
  year={2021},
  publisher={Nature Publishing Group UK London},
  url={https://doi.org/10.1038/s41567-020-1031-5}
}

@article{palomaki2013entangling,
  title={Entangling mechanical motion with microwave fields},
  author={Palomaki, TA and Teufel, JD and Simmonds, RW and Lehnert, Konrad W},
  journal={Science},
  volume={342},
  number={6159},
  pages={710--713},
  year={2013},
  publisher={American Association for the Advancement of Science},
  url={https://doi.org/10.1126/science.1244563}
}

@article{eisenberg2004quantum,
  title={Quantum entanglement of a large number of photons},
  author={Eisenberg, HS and Khoury, G and Durkin, GA and Simon, Christoph and Bouwmeester, D},
  journal={Physical Review Letters},
  volume={93},
  number={19},
  pages={193901},
  year={2004},
  publisher={APS},
  url={https://doi.org/10.1103/PhysRevLett.93.193901}
}

@article{afek2010high,
  title={High-NOON states by mixing quantum and classical light},
  author={Afek, Itai and Ambar, Oron and Silberberg, Yaron},
  journal={Science},
  volume={328},
  number={5980},
  pages={879--881},
  year={2010},
  publisher={American Association for the Advancement of Science},
  url={https://doi.org/10.1126/science.1188172}
}

@article{opatrny2000improvement,
  title={Improvement on teleportation of continuous variables by photon subtraction via conditional measurement},
  author={Opatrn{\`y}, T and Kurizki, Gershon and Welsch, D-G},
  journal={Physical Review A},
  volume={61},
  number={3},
  pages={032302},
  year={2000},
  publisher={APS},
  url={https://doi.org/10.1103/PhysRevA.61.032302}
}

@article{cochrane2002teleportation,
  title={Teleportation improvement by conditional measurements on the two-mode squeezed vacuum},
  author={Cochrane, PT and Ralph, TC and Milburn, GJ},
  journal={Physical Review A},
  volume={65},
  number={6},
  pages={062306},
  year={2002},
  publisher={APS},
  url={https://doi.org/10.1103/PhysRevA.65.062306}
}

@article{olivares2003teleportation,
  title={Teleportation improvement by inconclusive photon subtraction},
  author={Olivares, Stefano and Paris, Matteo GA and Bonifacio, Rodolfo},
  journal={Physical Review A},
  volume={67},
  number={3},
  pages={032314},
  year={2003},
  publisher={APS},
  url={https://doi.org/10.1103/PhysRevA.67.032314}
}

@article{johansson2012qutip,
  title={QuTiP: An open-source Python framework for the dynamics of open quantum systems},
  author={Johansson, J Robert and Nation, Paul D and Nori, Franco},
  journal={Computer Physics Communications},
  volume={183},
  number={8},
  pages={1760--1772},
  year={2012},
  publisher={Elsevier},
  url={https://doi.org/10.1016/j.cpc.2012.02.021}
}

@article{jeong2014generation,
  title={Generation of hybrid entanglement of light},
  author={Jeong, Hyunseok and Zavatta, Alessandro and Kang, Minsu and Lee, Seung-Woo and Costanzo, Luca S and Grandi, Samuele and Ralph, Timothy C and Bellini, Marco},
  journal={Nature Photonics},
  volume={8},
  number={7},
  pages={564--569},
  year={2014},
  publisher={Nature Publishing Group UK London},
  url={https://doi.org/10.1038/nphoton.2014.136}
}

@article{morin2014remote,
  title={Remote creation of hybrid entanglement between particle-like and wave-like optical qubits},
  author={Morin, Olivier and Huang, Kun and Liu, Jianli and Le Jeannic, Hanna and Fabre, Claude and Laurat, Julien},
  journal={Nature Photonics},
  volume={8},
  number={7},
  pages={570--574},
  year={2014},
  publisher={Nature Publishing Group UK London},
  url={https://doi.org/10.1038/nphoton.2014.137}
}

@article{ra2020non,
  title={Non-Gaussian quantum states of a multimode light field},
  author={Ra, Young-Sik and Dufour, Adrien and Walschaers, Mattia and Jacquard, Cl{\'e}ment and Michel, Thibault and Fabre, Claude and Treps, Nicolas},
  journal={Nature Physics},
  volume={16},
  number={2},
  pages={144--147},
  year={2020},
  publisher={Nature Publishing Group UK London},
  url={https://doi.org/10.1038/s41567-019-0726-y}
}

@article{xu2022metrological,
  title={Metrological characterization of non-gaussian entangled states of superconducting qubits},
  author={Xu, Kai and Zhang, Yu-Ran and Sun, Zheng-Hang and Li, Hekang and Song, Pengtao and Xiang, Zhongcheng and Huang, Kaixuan and Li, Hao and Shi, Yun-Hao and Chen, Chi-Tong and others},
  journal={Physical Review Letters},
  volume={128},
  number={15},
  pages={150501},
  year={2022},
  publisher={APS},
  url={https://doi.org/10.1103/PhysRevLett.128.150501}
}

@article{jost2009entangled,
  title={Entangled mechanical oscillators},
  author={Jost, John D and Home, JP and Amini, Jason M and Hanneke, David and Ozeri, Roee and Langer, Christopher and Bollinger, John J and Leibfried, Dietrich and Wineland, David J},
  journal={Nature},
  volume={459},
  number={7247},
  pages={683--685},
  year={2009},
  publisher={Nature Publishing Group UK London},
  url={https://doi.org/10.1038/nature08006}
}

@article{miki2022non,
  title={Non-Gaussian entanglement in gravitating masses: The role of cumulants},
  author={Miki, Daisuke and Matsumura, Akira and Yamamoto, Kazuhiro},
  journal={Physical Review D},
  volume={105},
  number={2},
  pages={026011},
  year={2022},
  publisher={APS},
  url={https://doi.org/10.1103/PhysRevD.105.026011}
}

@article{garcia2004proposal,
  title={Proposal for a loophole-free Bell test using homodyne detection},
  author={Garc{\'\i}a-Patr{\'o}n, Raul and Fiur{\'a}{\v{s}}ek, Jarom{\i}r and Cerf, Nicolas J and Wenger, J{\'e}r{\^o}me and Tualle-Brouri, Rosa and Grangier, Ph},
  journal={Physical Review Letters},
  volume={93},
  number={13},
  pages={130409},
  year={2004},
  publisher={APS},
  url={https://doi.org/10.1103/PhysRevLett.93.130409}
}

@article{dell2007continuous,
  title={Continuous-variable quantum teleportation with non-Gaussian resources},
  author={Dell’Anno, F and De Siena, S and Albano, L and Illuminati, F},
  journal={Physical Review A},
  volume={76},
  number={2},
  pages={022301},
  year={2007},
  publisher={APS},
  url={https://doi.org/10.1103/PhysRevA.76.022301}
}

@article{brunner2014bell,
  title={Bell nonlocality},
  author={Brunner, Nicolas and Cavalcanti, Daniel and Pironio, Stefano and Scarani, Valerio and Wehner, Stephanie},
  journal={Reviews of Modern Physics},
  volume={86},
  number={2},
  pages={419},
  year={2014},
  publisher={APS},
  url={https://doi.org/10.1103/RevModPhys.86.419}
}

@article{furry1936note,
  title={Note on the quantum-mechanical theory of measurement},
  author={Furry, Wendell H},
  journal={Physical Review},
  volume={49},
  number={5},
  pages={393},
  year={1936},
  publisher={APS},
  url={https://doi.org/10.1103/PhysRev.49.393}
}

@article{kiesewetter2017pulsed,
  title={Pulsed entanglement of two optomechanical oscillators and Furry’s hypothesis},
  author={Kiesewetter, S and Teh, RY and Drummond, PD and Reid, MD},
  journal={Physical Review Letters},
  volume={119},
  number={2},
  pages={023601},
  year={2017},
  publisher={APS},
  url={https://doi.org/10.1103/PhysRevLett.119.023601}
}

@article{patel2021room,
  title={Room-temperature mechanical resonator with a single added or subtracted phonon},
  author={Patel, Rishi N and McKenna, Timothy P and Wang, Zhaoyou and Witmer, Jeremy D and Jiang, Wentao and Van Laer, Rapha{\"e}l and Sarabalis, Christopher J and Safavi-Naeini, Amir H},
  journal={Physical Review Letters},
  volume={127},
  number={13},
  pages={133602},
  year={2021},
  publisher={APS},
  url={https://doi.org/10.1103/PhysRevLett.127.133602}
}

@article{akram2013entangled,
  title={Entangled mechanical cat states via conditional single photon optomechanics},
  author={Akram, Uzma and Bowen, Warwick P and Milburn, Gerard J},
  journal={New Journal of Physics},
  volume={15},
  number={9},
  pages={093007},
  year={2013},
  publisher={IOP Publishing},
  url={https://doi.org/10.1088/1367-2630/15/9/093007}
}

@article{lloyd1999quantum,
  title={Quantum computation over continuous variables},
  author={Lloyd, Seth and Braunstein, Samuel L},
  journal={Physical Review Letters},
  volume={82},
  number={8},
  pages={1784},
  year={1999},
  publisher={APS},
  url={https://doi.org/10.1103/PhysRevLett.82.1784}
}

@article{gottesman2001encoding,
  title={Encoding a qubit in an oscillator},
  author={Gottesman, Daniel and Kitaev, Alexei and Preskill, John},
  journal={Physical Review A},
  volume={64},
  number={1},
  pages={012310},
  year={2001},
  publisher={APS},
  url={https://doi.org/10.1103/PhysRevA.64.012310}
}

@book{meystre2021quantum,
  title={Quantum optics},
  author={Meystre, Pierre},
  year={2021},
  publisher={Springer},
  url={https://www.google.co.uk/books/edition/Elements_of_Quantum_Optics/81GSjqCIIFAC?hl=en&gbpv=0}
}

@article{kanari2021can,
  title={Can the displacemon device test objective collapse models?},
  author={Kanari-Naish, Lydia A and Clarke, Jack and Vanner, Michael R and Laird, Edward A},
  journal={AVS Quantum Science},
  volume={3},
  number={4},
  pages={045603},
  year={2021},
  publisher={American Vacuum Society},
  url={https://doi.org/10.1116/5.0073626}
}

@article{clarke2018growing,
  title={Growing macroscopic superposition states via cavity quantum optomechanics},
  author={Clarke, Jack and Vanner, Michael R},
  journal={Quantum Science and Technology},
  volume={4},
  number={1},
  pages={014003},
  year={2019},
  publisher={IOP Publishing},
  url={https://doi.org/10.1088/2058-9565/aada1d}
}

@article{gilchrist2004schrodinger,
  title={Schr{\"o}dinger cats and their power for quantum information processing},
  author={Gilchrist, Alexei and Nemoto, Kae and Munro, William J and Ralph, Timothy C and Glancy, Scott and Braunstein, Samuel L and Milburn, Gerard J},
  journal={Journal of Optics B: Quantum and Semiclassical Optics},
  volume={6},
  number={8},
  pages={S828},
  year={2004},
  publisher={IOP Publishing},
  url={https://doi.org/10.1088/1464-4266/6/8/032}
}

@article{giovannetti2011advances,
  title={Advances in quantum metrology},
  author={Giovannetti, Vittorio and Lloyd, Seth and Maccone, Lorenzo},
  journal={Nature Photonics},
  volume={5},
  number={4},
  pages={222--229},
  year={2011},
  publisher={Nature Publishing Group UK London},
  url={https://doi.org/10.1038/nphoton.2011.35}
}

@article{jeong2002efficient,
  title={Efficient quantum computation using coherent states},
  author={Jeong, H and Kim, Myung Shik},
  journal={Physical Review A},
  volume={65},
  number={4},
  pages={042305},
  year={2002},
  publisher={APS},
  url={https://doi.org/10.1103/PhysRevA.65.042305}
}

@article{sanders2012review,
  title={Review of entangled coherent states},
  author={Sanders, Barry C},
  journal={Journal of Physics A: Mathematical and theoretical},
  volume={45},
  number={24},
  pages={244002},
  year={2012},
  publisher={IOP Publishing},
  url={https://doi.org/10.1088/1751-8113/45/24/244002}
}

@article{gao2019entanglement,
  title={Entanglement of bosonic modes through an engineered exchange interaction},
  author={Gao, Yvonne Y and Lester, Brian J and Chou, Kevin S and Frunzio, Luigi and Devoret, Michel H and Jiang, Liang and Girvin, SM and Schoelkopf, Robert J},
  journal={Nature},
  volume={566},
  number={7745},
  pages={509--512},
  year={2019},
  publisher={Nature Publishing Group UK London},
  url={https://doi.org/10.1038/s41586-019-0970-4}
}

@article{wang2022flying,
  title={A flying Schr{\"o}dinger’s cat in multipartite entangled states},
  author={Wang, Zhiling and Bao, Zenghui and Wu, Yukai and Li, Yan and Cai, Weizhou and Wang, Weiting and Ma, Yuwei and Cai, Tianqi and Han, Xiyue and Wang, Jiahui and others},
  journal={Science Advances},
  volume={8},
  number={10},
  pages={eabn1778},
  year={2022},
  publisher={American Association for the Advancement of Science},
  url={https://doi.org/10.1126/sciadv.abn1778}
}

@article{munro2000entangled,
  title={Entangled coherent-state qubits in an ion trap},
  author={Munro, WJ and Milburn, Gerard J and Sanders, BC},
  journal={Physical Review A},
  volume={62},
  number={5},
  pages={052108},
  year={2000},
  publisher={APS},
  url={https://doi.org/10.1103/PhysRevA.62.052108}
}

@article{moehring2007entanglement,
  title={Entanglement of single-atom quantum bits at a distance},
  author={Moehring, David L and Maunz, Peter and Olmschenk, Steve and Younge, Kelly C and Matsukevich, Dzmitry N and Duan, L-M and Monroe, Christopher},
  journal={Nature},
  volume={449},
  number={7158},
  pages={68--71},
  year={2007},
  publisher={Nature Publishing Group UK London},
  url={https://doi.org/10.1038/nature06118}
}

@article{song201710,
  title={10-qubit entanglement and parallel logic operations with a superconducting circuit},
  author={Song, Chao and Xu, Kai and Liu, Wuxin and Yang, Chui-ping and Zheng, Shi-Biao and Deng, Hui and Xie, Qiwei and Huang, Keqiang and Guo, Qiujiang and Zhang, Libo and others},
  journal={Physical Review Letters},
  volume={119},
  number={18},
  pages={180511},
  year={2017},
  publisher={APS},
  url={https://doi.org/10.1103/PhysRevLett.119.180511}
}

@article{adesso2009experimentally,
  title={Experimentally friendly bounds on non-Gaussian entanglement from second moments},
  author={Adesso, Gerardo},
  journal={Physical Review A},
  volume={79},
  number={2},
  pages={022315},
  year={2009},
  publisher={APS},
  url={https://doi.org/10.1103/PhysRevA.79.022315}
}

@article{zhang2010distillation,
  title={Distillation of mixed-state continuous-variable entanglement by photon subtraction},
  author={Zhang, Sheng Li and van Loock, Peter},
  journal={Physical Review A},
  volume={82},
  number={6},
  pages={062316},
  year={2010},
  publisher={APS},
  url={https://doi.org/10.1103/PhysRevA.82.062316}
}

@article{agarwal2005inseparability,
  title={Inseparability inequalities for higher order moments for bipartite systems},
  author={Agarwal, Girish Saran and Biswas, Asoka},
  journal={New Journal of Physics},
  volume={7},
  number={1},
  pages={211},
  year={2005},
  publisher={IOP Publishing},
  url={https://doi.org/10.1088/1367-2630/7/1/211}
}

@article{bose2023massive,
  title={Massive quantum systems as interfaces of quantum mechanics and gravity},
  author={Bose, Sougato and Fuentes, Ivette and Geraci, Andrew A and Khan, Saba Mehsar and Qvarfort, Sofia and Rademacher, Markus and Rashid, Muddassar and Toro{\v{s}}, Marko and Ulbricht, Hendrik and Wanjura, Clara C},
  journal={Reviews of Modern Physics},
  volume={97},
  number={1},
  pages={015003},
  year={2025},
  publisher={APS},
  url={https://doi.org/10.1103/RevModPhys.97.015003}
}

@article{weedbrook2012gaussian,
  title={Gaussian quantum information},
  author={Weedbrook, Christian and Pirandola, Stefano and Garc{\'\i}a-Patr{\'o}n, Ra{\'u}l and Cerf, Nicolas J and Ralph, Timothy C and Shapiro, Jeffrey H and Lloyd, Seth},
  journal={Reviews of Modern Physics},
  volume={84},
  number={2},
  pages={621},
  year={2012},
  publisher={APS},
  url={https://doi.org/10.1103/RevModPhys.84.621}
}

@article{furusawa1998unconditional,
  title={Unconditional quantum teleportation},
  author={Furusawa, Akira and S{\o}rensen, Jens Lykke and Braunstein, Samuel L and Fuchs, Christopher A and Kimble, H Jeff and Polzik, Eugene S},
  journal={Science},
  volume={282},
  number={5389},
  pages={706--709},
  year={1998},
  publisher={American Association for the Advancement of Science},
  url={https://doi.org/10.1126/science.282.5389.706}
}

@article{ou1992realization,
  title={Realization of the Einstein-Podolsky-Rosen paradox for continuous variables},
  author={Ou, ZY and Pereira, Silvana F and Kimble, HJ and Peng, KC},
  journal={Physical Review Letters},
  volume={68},
  number={25},
  pages={3663},
  year={1992},
  publisher={APS},
  url={https://doi.org/10.1103/PhysRevLett.68.3663}
}

@article{eberle2013stable,
  title={Stable control of 10 dB two-mode squeezed vacuum states of light},
  author={Eberle, Tobias and H{\"a}ndchen, Vitus and Schnabel, Roman},
  journal={Optics Express},
  volume={21},
  number={9},
  pages={11546--11553},
  year={2013},
  publisher={Optica Publishing Group},
  url={https://doi.org/10.1364/OE.21.011546}
}

@article{anisimov2010quantum,
  title={Quantum metrology with two-mode squeezed vacuum: parity detection beats the Heisenberg limit},
  author={Anisimov, Petr M and Raterman, Gretchen M and Chiruvelli, Aravind and Plick, William N and Huver, Sean D and Lee, Hwang and Dowling, Jonathan P},
  journal={Physical Review Letters},
  volume={104},
  number={10},
  pages={103602},
  year={2010},
  publisher={APS},
  url={https://doi.org/10.1103/PhysRevLett.104.103602}
}

@article{ou1997fundamental,
  title={Fundamental quantum limit in precision phase measurement},
  author={Ou, ZY},
  journal={Physical Review A},
  volume={55},
  number={4},
  pages={2598},
  year={1997},
  publisher={APS},
  url={https://doi.org/10.1103/PhysRevA.55.2598}
}

@article{grosshans2003quantum,
  title={Quantum key distribution using gaussian-modulated coherent states},
  author={Grosshans, Fr{\'e}d{\'e}ric and Van Assche, Gilles and Wenger, J{\'e}r{\^o}me and Brouri, Rosa and Cerf, Nicolas J and Grangier, Philippe},
  journal={Nature},
  volume={421},
  number={6920},
  pages={238--241},
  year={2003},
  publisher={Nature Publishing Group UK London},
  url={https://doi.org/10.1038/nature01289}
}

@article{braunstein2005quantum,
  title={Quantum information with continuous variables},
  author={Braunstein, Samuel L and Van Loock, Peter},
  journal={Reviews of Modern Physics},
  volume={77},
  number={2},
  pages={513},
  year={2005},
  publisher={APS},
  url={https://doi.org/10.1103/RevModPhys.77.513}
}

@article{zavatta2004quantum,
  title={Quantum-to-classical transition with single-photon-added coherent states of light},
  author={Zavatta, Alessandro and Viciani, Silvia and Bellini, Marco},
  journal={Science},
  volume={306},
  number={5696},
  pages={660--662},
  year={2004},
  publisher={American Association for the Advancement of Science},
  url={https://doi.org/10.1126/science.1103190}
}

@article{zavatta2007experimental,
  title={Experimental nonclassicality of single-photon-added thermal light states},
  author={Zavatta, Alessandro and Parigi, Valentina and Bellini, Marco},
  journal={Physical Review A},
  volume={75},
  number={5},
  pages={052106},
  year={2007},
  publisher={APS},
  url={https://doi.org/10.1103/PhysRevA.75.052106}
}

@article{qvarfort2020mesoscopic,
  title={Mesoscopic entanglement through central--potential interactions},
  author={Qvarfort, Sofia and Bose, Sougato and Serafini, Alessio},
  journal={Journal of Physics B: Atomic, Molecular and Optical Physics},
  volume={53},
  number={23},
  pages={235501},
  year={2020},
  publisher={IOP Publishing},
  doi={10.1088/1361-6455/abbe8d}
}

@article{takahashi2010entanglement,
  title={Entanglement distillation from Gaussian input states},
  author={Takahashi, Hiroki and Neergaard-Nielsen, Jonas S and Takeuchi, Makoto and Takeoka, Masahiro and Hayasaka, Kazuhiro and Furusawa, Akira and Sasaki, Masahide},
  journal={Nature Photonics},
  volume={4},
  number={3},
  pages={178--181},
  year={2010},
  publisher={Nature Publishing Group UK London},
  url={https://doi.org/10.1038/nphoton.2010.1}
}

@article{ourjoumtsev2007increasing,
  title={Increasing entanglement between Gaussian states by coherent photon subtraction},
  author={Ourjoumtsev, Alexei and Dantan, Aurelien and Tualle-Brouri, Rosa and Grangier, Philippe},
  journal={Physical Review Letters},
  volume={98},
  number={3},
  pages={030502},
  year={2007},
  publisher={APS},
  url={https://doi.org/10.1103/PhysRevLett.98.030502}
}

@article{hong1999statistical,
  title={Statistical properties of photon-added and photon-subtracted two-mode squeezed vacuum state},
  author={Hong, Lu},
  journal={Physics Letters A},
  volume={264},
  number={4},
  pages={265--269},
  year={1999},
  publisher={Elsevier},
  url={https://doi.org/10.1016/S0375-9601(99)00802-6}
}

@article{rauschenbeutel2001controlled,
  title={Controlled entanglement of two field modes in a cavity quantum electrodynamics experiment},
  author={Rauschenbeutel, A and Bertet, Patrice and Osnaghi, S and Nogues, Gilles and Brune, Michel and Raimond, Jean-Michel and Haroche, Serge},
  journal={Physical Review A},
  volume={64},
  number={5},
  pages={050301},
  year={2001},
  publisher={APS},
  url={https://doi.org/10.1103/PhysRevA.64.050301}
}

@article{wang2011deterministic,
  title={Deterministic entanglement of photons in two superconducting microwave resonators},
  author={Wang, H and Mariantoni, Matteo and Bialczak, Radoslaw C and Lenander, M and Lucero, Erik and Neeley, M and O’Connell, AD and Sank, D and Weides, M and Wenner, J and others},
  journal={Physical Review Letters},
  volume={106},
  number={6},
  pages={060401},
  year={2011},
  publisher={APS},
  url={https://doi.org/10.1103/PhysRevLett.106.060401}
}

@article{zhang2018noon,
  title={NOON states of nine quantized vibrations in two radial modes of a trapped ion},
  author={Zhang, Junhua and Um, Mark and Lv, Dingshun and Zhang, Jing-Ning and Duan, Lu-Ming and Kim, Kihwan},
  journal={Physical Review Letters},
  volume={121},
  number={16},
  pages={160502},
  year={2018},
  publisher={APS},
  url={https://doi.org/10.1103/PhysRevLett.121.160502}
}

@article{jeon2024experimental,
  title={Experimental realization of entangled coherent states in two-dimensional harmonic oscillators of a trapped ion},
  author={Jeon, Honggi and Kang, Jiyong and Kim, Jaeun and Choi, Wonhyeong and Kim, Kyunghye and Kim, Taehyun},
  journal={Scientific Reports},
  volume={14},
  number={1},
  pages={6847},
  year={2024},
  publisher={Nature Publishing Group UK London},
  url={https://doi.org/10.1038/s41598-024-57391-6}
}

@article{bowen2003experimental,
  title={Experimental investigation of criteria for continuous variable entanglement},
  author={Bowen, Warwick P and Schnabel, Raik and Lam, Ping Koy and Ralph, Timothy Cameron},
  journal={Physical Review Letters},
  volume={90},
  number={4},
  pages={043601},
  year={2003},
  publisher={APS},
  url={https://doi.org/10.1103/PhysRevLett.90.043601}
}

@article{plato2023enhanced,
  title={Enhanced Gravitational Entanglement via Modulated Optomechanics},
  author={Plato, A Douglas K and R{\"a}tzel, Dennis and Wan, Chuanqi},
  journal={Quantum},
  volume={7},
  pages={1177},
  year={2023},
  publisher={Verein zur F{\"o}rderung des Open Access Publizierens in den Quantenwissenschaften},
  url={https://doi.org/10.22331/q-2023-11-08-1177}
}

@article{clarke2020generating,
  title={Generating mechanical and optomechanical entanglement via pulsed interaction and measurement},
  author={Clarke, J and Sahium, P and Khosla, K E and Pikovski, I and Kim, M S and Vanner, M R},
  journal={New Journal of Physics},
  volume={22},
  number={6},
  pages={063001},
  year={2020},
  publisher={IOP Publishing},
  url={https://doi.org/10.1088/1367-2630/ab7ddd}
}

@article{neveu2021preparation,
  title={Preparation and verification of two-mode mechanical entanglement through pulsed optomechanical measurements},
  author={Neveu, Pascal and Clarke, Jack and Vanner, Michael R and Verhagen, Ewold},
  journal={New Journal of Physics},
  volume={23},
  number={2},
  pages={023026},
  year={2021},
  publisher={IOP Publishing},
  url={https://doi.org/10.1088/1367-2630/abe1e4}
}

@article{vanner2011pulsed,
  title={Pulsed quantum optomechanics},
  author={Vanner, Michael R and Pikovski, Igor and Cole, Garrett D and Kim, Myung Shik and Brukner, {\v{C}} and Hammerer, Klemens and Milburn, Gerard J and Aspelmeyer, Markus},
  journal={Proceedings of the National Academy of Sciences},
  volume={108},
  number={39},
  pages={16182--16187},
  year={2011},
  publisher={National Acad Sciences},
  url={https://doi.org/10.1073/pnas.1105098108}
}

@article{vanner2013cooling,
  title={Cooling-by-measurement and mechanical state tomography via pulsed optomechanics},
  author={Vanner, M R and Hofer, J and Cole, G D and Aspelmeyer, M},
  journal={Nature Communications},
  volume={4},
  number={1},
  pages={2295},
  year={2013},
  publisher={Nature Publishing Group UK London},
  url={https://doi.org/10.1038/ncomms3295}
}

@article{muhonen2019state,
  title={State preparation and tomography of a nanomechanical resonator with fast light pulses},
  author={Muhonen, Juha T and La Gala, Giada R and Leijssen, Rick and Verhagen, Ewold},
  journal={Physical Review Letters},
  volume={123},
  number={11},
  pages={113601},
  year={2019},
  publisher={APS},
  url={https://doi.org/10.1103/PhysRevLett.123.113601}
}

@article{tan1999confirming,
  title={Confirming entanglement in continuous variable quantum teleportation},
  author={Tan, Sze M},
  journal={Physical Review A},
  volume={60},
  number={4},
  pages={2752},
  year={1999},
  publisher={APS},
  url={https://doi.org/10.1103/PhysRevA.60.2752}
}

@misc{lydia_code,
    note={\url{https://github.com/QuantumMeasurementLab/NPyT}}
}

@article{guhne2009entanglement,
  title={Entanglement detection},
  author={G{\"u}hne, Otfried and T{\'o}th, G{\'e}za},
  journal={Physics Reports},
  volume={474},
  number={1-6},
  pages={1--75},
  year={2009},
  publisher={Elsevier},
  url={https://doi.org/10.1016/j.physrep.2009.02.004}
}

@article{tavakoli2024quantum,
  title={Quantum steering with imprecise measurements},
  author={Tavakoli, Armin},
  journal={Physical Review Letters},
  volume={132},
  number={7},
  pages={070204},
  year={2024},
  publisher={APS},
  url={https://doi.org/10.1103/PhysRevLett.132.070204}
}

@article{cao2024genuine,
  title={Genuine Multipartite Entanglement Detection with Imperfect Measurements: Concept and Experiment},
  author={Cao, Huan and Morelli, Simon and Rozema, Lee A and Zhang, Chao and Tavakoli, Armin and Walther, Philip},
  journal={Physical Review Letters},
  volume={133},
  number={15},
  pages={150201},
  year={2024},
  publisher={APS},
  url={https://doi.org/10.1103/PhysRevLett.133.150201}
}

@article{peres1996separability,
  title={Separability criterion for density matrices},
  author={Peres, Asher},
  journal={Physical Review Letters},
  volume={77},
  number={8},
  pages={1413},
  year={1996},
  publisher={APS},
  url={https://doi.org/10.1103/PhysRevLett.77.1413}
}

@article{nielsen2017multimode,
author = {William Hvidtfelt Padkær Nielsen  and Yeghishe Tsaturyan  and Christoffer Bo Møller  and Eugene S. Polzik  and Albert Schliesser },
title = {Multimode optomechanical system in the quantum regime},
journal = {Proceedings of the National Academy of Sciences},
volume = {114},
number = {1},
pages = {62-66},
year = {2017},
URL = {https://www.pnas.org/doi/abs/10.1073/pnas.1608412114}
}

@article{vanner2011selective,
  title = {Selective Linear or Quadratic Optomechanical Coupling via Measurement},
  author = {Vanner, Michael R.},
  journal = {Phys. Rev. X},
  volume = {1},
  issue = {2},
  pages = {021011},
  numpages = {7},
  year = {2011},
  publisher = {American Physical Society},
  url = {https://link.aps.org/doi/10.1103/PhysRevX.1.021011}
}

@article{brawley2016nonlinear,
  title={Nonlinear optomechanical measurement of mechanical motion},
  author={Brawley, G A and Vanner, M R and Larsen, P E and Schmid, S and Boisen, A and Bowen, W P},
  journal={Nature Communications},
  volume={7},
  number={1},
  pages={10988},
  year={2016},
  url={http://www.nature.com/ncomms/2016/160321/ncomms10988/full/ncomms10988.html}
}

@article{leijssen2017nonlinear,
  title={Nonlinear cavity optomechanics with nanomechanical thermal fluctuations},
  author={Leijssen, Rick and La Gala, Giada R and Freisem, Lars and Muhonen, Juha T and Verhagen, Ewold},
  journal={Nature Communications},
  volume={8},
  number={1},
  pages={16024},
  year={2017},
  url={https://www.nature.com/articles/ncomms16024}
}

@article{clarke2023cavity,
  title={Cavity quantum optomechanical nonlinearities and position measurement beyond the breakdown of the linearized approximation},
  author={Clarke, Jack and Neveu, Pascal and Khosla, Kiran E and Verhagen, Ewold and Vanner, Michael R},
  journal={Physical Review Letters},
  volume={131},
  number={5},
  pages={053601},
  year={2023},
  url={https://journals.aps.org/prl/abstract/10.1103/PhysRevLett.131.053601}
}

@article{miranowicz2010testing,
  title={Testing nonclassicality in multimode fields: A unified derivation of classical inequalities},
  author={Miranowicz, Adam and Bartkowiak, Monika and Wang, Xiaoguang and Liu, Yu-xi and Nori, Franco},
  journal={Physical Review A},
  volume={82},
  number={1},
  pages={013824},
  year={2010},
  publisher={APS},
  url={https://doi.org/10.1103/PhysRevA.82.013824}
}

@article{xia2014opto,
  title={An opto-magneto-mechanical quantum interface between distant superconducting qubits},
  author={Xia, Keyu and Vanner, Michael R and Twamley, Jason},
  journal={Scientific Reports},
  volume={4},
  number={1},
  pages={5571},
  year={2014},
  publisher={Nature Publishing Group UK London},
  url={https://www.nature.com/articles/srep05571}
}

@article{patil2022measuring,
  title={Measuring high-order phonon correlations in an optomechanical resonator},
  author={Patil, Yogesh SS and Yu, Jiaxin and Frazier, Sean and Wang, Yiqi and Johnson, Kale and Fox, Jared and Reichel, Jakob and Harris, Jack GE},
  journal={Physical Review Letters},
  volume={128},
  number={18},
  pages={183601},
  year={2022},
  publisher={APS},
  url={https://journals.aps.org/prl/abstract/10.1103/PhysRevLett.128.183601}
}

@article{cryer2025enhanced,
  title={Enhanced laser cooling of a mechanical resonator via zero-photon detection},
  author={Cryer-Jenkins, Evan A and Major, Kyle D and Clarke, Jack and Enzian, Georg and Szczykulska, Magdalena and Zhang, Jinglei and Gupta, Arjun and Leung, Anthony C and Rathee, Harsh and Svela, Andreas {\O} and others},
  journal={Physical Review Letters},
  volume={134},
  number={7},
  pages={073601},
  year={2025},
  publisher={APS},
  url={https://journals.aps.org/prl/abstract/10.1103/PhysRevLett.134.073601}
}

@article{clarke2025theoretical,
  title={Theoretical framework for enhancing or enabling cooling of a mechanical resonator via the anti-Stokes or Stokes interaction and zero-photon detection},
  author={Clarke, Jack and Cryer-Jenkins, Evan A and Gupta, Arjun and Major, Kyle D and Zhang, Jinglei and Enzian, Georg and Szczykulska, Magdalena and Leung, Anthony C and Rathee, Harsh and Svela, Andreas {\O} and others},
  journal={Physical Review A},
  volume={111},
  number={2},
  pages={023516},
  year={2025},
  publisher={APS},
  url={https://journals.aps.org/pra/abstract/10.1103/PhysRevA.111.023516}
}

@article{horodecki1998mixed,
  title={Mixed-state entanglement and distillation: Is there a “bound” entanglement in nature?},
  author={Horodecki, Micha{\l} and Horodecki, Pawe{\l} and Horodecki, Ryszard},
  journal={Physical Review Letters},
  volume={80},
  number={24},
  pages={5239},
  year={1998},
  publisher={APS},
  url={https://journals.aps.org/prl/abstract/10.1103/PhysRevLett.80.5239}
}

\appendix
\begin{widetext}
\onecolumngrid

\section{Entanglement criteria and ordering of multi-indices}
\label{app:entanglement_criteria}

The matrix $\mathbf{M}$ has entries given by Eq.~\eqref{eq:super_matrix_ij} and the matrix indices $i$ and $j$ are the ordinal numbers of two sets of multi-indices. The precise ordering is arbitrary, but once it is picked it must be kept consistent across the two sets of multi-indices. Here, the first few ordered multi-indices according to the ordering we have adopted are listed below in Table~\ref{tab:multi_index}.

\begin{table}[h]
\caption{The ordering convention we have adopted to generate the indices of the matrix $\mathbf{M}$ defined in Eq.~\eqref{eq:super_matrix_ij}. We have listed $i,j\leq 15$ as these indices are sufficient to specify all moments up to order $n=4$, that is,} any submatrix $\mathrm{A}$ generated by keeping rows and columns of $\mathbf{M}$ up to $i,j\leq15$ will have $n\leq 4$. If the indices $i,j>15$ are included, higher order moments will be generated.
\begingroup
\setlength{\tabcolsep}{1pt}
\begin{ruledtabular}
\begin{tabular}{cccc}
$i$ & $(p_i,q_i,r_i,s_i)$  & $j$ & $(n_j,m_j,k_j,l_j)$ \\
  \hline
  1 & (0, 0, 0, 0) & 1 & (0, 0, 0, 0) \\
  2 & (1, 0, 0, 0) & 2 & (1, 0, 0, 0) \\
  3 & (0, 1, 0, 0) & 3 & (0, 1, 0, 0) \\
  4 & (0, 0, 1, 0) & 4 & (0, 0, 1, 0) \\
  5 & (0, 0, 0, 1) & 5 & (0, 0, 0, 1) \\
  6 & (2, 0, 0, 0) & 6 & (2, 0, 0, 0) \\
  7 &  (1, 1, 0, 0) & 7 &  (1, 1, 0, 0) \\
  8 & (0, 2, 0, 0) & 8 & (0, 2, 0, 0) \\
  9 & (1, 0, 1, 0) & 9 & (1, 0, 1, 0) \\
  10 & (0, 1, 1, 0) & 10 & (0, 1, 1, 0) \\
  11 & (0, 0, 2, 0) & 11 & (0, 0, 2, 0) \\
  12 & (1, 0, 0, 1) & 12 & (1, 0, 0, 1) \\
  13 & (0, 1, 0, 1) & 13 & (0, 1, 0, 1) \\
  14 & (0, 0, 1, 1) & 14 & (0, 0, 1, 1) \\
  15 & (0, 0, 0, 2) & 15 & (0, 0, 0, 2)\\
\end{tabular}
\end{ruledtabular}
\endgroup
\label{tab:multi_index}
\end{table}

Using the ordering in Table~\ref{tab:multi_index}, the matrix $\mathbf{M}$ is
\begin{equation}
\label{eq:super_matrix}
    \mathbf{M}=\begin{pmatrix}
        1 & \braket{\hat{a}^\dag} & \braket{\hat{a}} & \braket{\hat{b}} & \braket{\hat{b}^\dag} & \braket{\hat{a}^{\dag 2}} & \braket{\hat{a}^\dag \hat{a}} &  \ldots \\
       \braket{\hat{a}} & \braket{\hat{a}\hat{a}^\dag} & \braket{\hat{a}^2} & \braket{\hat{a}\hat{b}} & \braket{\hat{a}\hat{b}^\dag} & \braket{\hat{a}\hat{a}^{\dag 2}} &  \braket{\hat{a}\hat{a}^\dag \hat{a}} & \ldots \\
       \braket{\hat{a}^\dag} & \braket{\hat{a}^{\dag 2} } & \braket{\hat{a}^\dag \hat{a}} & \braket{\hat{a}^\dag\hat{b}} & \braket{\hat{a}^\dag\hat{b}^\dag} & \braket{\hat{a}^{\dag 3}} & \braket{\hat{a}^{\dag 2}\hat{a}} & \ldots \\
       \braket{\hat{b}^\dag} & \braket{\hat{a}^\dag \hat{b}^\dag} & \braket{\hat{a}\hat{b}^\dag} & \braket{\hat{b}\hat{b}^\dag} & \braket{\hat{b}^{\dag 2}} & \braket{\hat{a}^{\dag 2}\hat{b}^\dag} & \braket{\hat{a}^\dag \hat{b}\hat{b}^\dag} & \ldots \\
       \braket{\hat{b}} & \braket{\hat{a}^\dag\hat{b}} & \braket{\hat{a}\hat{b}} & \braket{\hat{b}^2} & \braket{\hat{b}^\dag\hat{b}} & \braket{\hat{a}^{\dag 2}\hat{b}} & \braket{\hat{a}^\dag \hat{a}\hat{b}} & \ldots \\
       \braket{\hat{a}^2} & \braket{\hat{a}^2 \hat{a}^\dag} & \braket{\hat{a}^3} & \braket{\hat{a}^2\hat{b}} & \braket{\hat{a}^2\hat{b}^\dag} & \braket{\hat{a}^2\hat{a}^{\dag 2}} & \braket{\hat{a}^2 \hat{a}^\dag \hat{a}}  & \ldots\\
       \braket{\hat{a}^\dag \hat{a}} & \braket{\hat{a}^\dag \hat{a}\hat{a}^\dag} & \braket{\hat{a}^\dag \hat{a}^2} & \braket{\hat{a}^\dag \hat{a}\hat{b}} & \braket{\hat{a}^\dag \hat{a}\hat{b}^\dag} & \braket{\hat{a}^\dag \hat{a}\hat{a}^2} & \braket{\hat{a}^\dag \hat{a}\hat{a}^\dag\hat{a}} & \ldots \\
       \vdots & \vdots & \vdots & \vdots & \vdots & \vdots & \vdots & \ddots
    \end{pmatrix}~.
\end{equation}
Taking just the first 5 rows and columns leaves the leading principal minor $\mathbf{M}_{5}$, which gives a necessary and sufficient criterion for Gaussian states via Simon's criterion: all entangled Gaussian states will have $\mathrm{det}[\mathbf{M}_5]<0$. We can also generate other criteria by deleting rows and columns of $\mathbf{M}$ in a pairwise fashion to leave behind a $d\times d$ submatrix $\mathbf{A}$. If $\mathrm{det}[\mathbf{A}]<0$ can be shown, then entanglement is present. 
In Table~\ref{tab:multi_index}, we have only listed $i,j\leq 15$ as these indices are sufficient to specify any submatrix $\mathbf{A}$ with $n\leq 4$. 
Our statistical framework is applicable for arbitrarily high $i$ and $j$ but this is at the cost of increased experimental complexity and sampling errors. For the purposes of demonstrating our statistical framework, we have limited our discussion here to the case $n\leq 4$. 

We identify specific NPT criteria that are able to detect entanglement in three example states: (i) the TMSV, (ii) the single-quantum-subtracted TMSV state, and (iii) the two-mode Schr{\"o}dinger-cat state. For the parameters explored in the main text, the NPT criteria that are successful at detecting entanglement are listed in Table~\ref{tab:TMSV}, Table~\ref{tab:TMSV_sub}, and Table~\ref{tab:TMSC} for the TMSV, the single-quantum-subtracted TMSV state, and the two-mode Schr{\"o}dinger-cat state, respectively.

\begin{table}[h]
\caption{Suitable NPT criteria for the TMSV state $\ket{\Psi}_{\mathrm{TMSV}}$.
The matrix $\mathbf{A}$ from which the determinant is calculated is provided, along with the dimension $d$, the order $n$ of the highest-order moment, and the rows/columns that are kept from the matrix $\mathbf{M}$.
Determinants $D_{\mathrm{I-VIII}}$ are the only determinants with $d\leq 5$ and $n\leq 2$ that exhibit negativity in the region of parameter space explored in Fig.~\ref{fig:results_TMSV}. The indices of the rows and columns that are kept in order to generate $\mathbf{A}$ are shown. For this state, $D_{\mathrm{I}}=D_{\mathrm{IV}}$, $D_{\mathrm{II}}=D_{\mathrm{V}}=D_{\mathrm{VII}}=D_{\mathrm{VIII}}$, and $D_{\mathrm{III}}=D_{\mathrm{VI}}$.
}
\begin{ruledtabular}
\begin{tabular}{l c c c r}
Determinant & Matrix form, $\mathrm{det}[\mathbf{A}]$  & Dimension, $d$ & Order, $n$ & Rows/Columns  \\
  \hline
  $D_{\mathrm{I}}$ & $\begin{vmatrix}
  \braket{\hat{a}^\dag \hat{a}} & \braket{\hat{a}^\dag \hat{b}^{\dag}}\\
  \braket{\hat{a}\hat{b}} & \braket{\hat{b}^\dag \hat{b}}
\end{vmatrix}$ & 2 & 2 & (3,5) \\

  $D_\mathrm{II}$ & $\begin{vmatrix}
  \braket{\hat{a} \hat{a}^{\dag}} & \braket{\hat{a}\hat{a}} & \braket{\hat{a} \hat{b}^{\dag}} \\
  \braket{\hat{a}^{\dag} \hat{a}^{\dag}} & \braket{\hat{a}^\dag \hat{a}} & \braket{\hat{a}^\dag \hat{b}^{\dag}}\\
  \braket{\hat{a}^{\dag} \hat{b}} & \braket{\hat{a}\hat{b}} & \braket{\hat{b}^\dag \hat{b}}
\end{vmatrix}$ & 3 & 2 & (2,3,5) \\

$D_\mathrm{III}$ & $\begin{vmatrix}
\braket{\hat{a}\hat{a}^\dag} & \braket{\hat{a}^2} & \braket{\hat{a}\hat{b}} & \braket{\hat{a}\hat{b}^\dag} \\
\braket{\hat{a}^{\dag 2} } & \braket{\hat{a}^\dag \hat{a}} & \braket{\hat{a}^\dag\hat{b}} & \braket{\hat{a}^\dag\hat{b}^\dag} \\
 \braket{\hat{a}^\dag \hat{b}^\dag} & \braket{\hat{a}\hat{b}^\dag} & \braket{\hat{b}\hat{b}^\dag} & \braket{\hat{b}^{\dag 2}} \\
  \braket{\hat{a}^\dag\hat{b}} & \braket{\hat{a}\hat{b}} & \braket{\hat{b}^2} & \braket{\hat{b}^\dag\hat{b}} \\
\end{vmatrix}$ & 4 & 2 & (2,3,4,5) \\

  $D_\mathrm{IV}$ & $\begin{vmatrix}
  1 & \braket{\hat{a}} & \braket{\hat{b}^{\dag}} \\
  \braket{\hat{a}^{\dag}} & \braket{\hat{a}^\dag \hat{a}} & \braket{\hat{a}^\dag \hat{b}^{\dag}}\\
  \braket{\hat{b}} & \braket{\hat{a}\hat{b}} & \braket{\hat{b}^\dag \hat{b}}
\end{vmatrix}$ & 3 & 2 & (1,3,5) \\

  $D_\mathrm{V}$ & $\begin{vmatrix}
  1 & \braket{\hat{a}^{\dag}} & \braket{\hat{a}} & \braket{\hat{b}} \\
  \braket{\hat{a}} & \braket{\hat{a} \hat{a}^{\dag}} & \braket{\hat{a}\hat{a}} & \braket{\hat{a} \hat{b}^{\dag}} \\
  \braket{\hat{a}^{\dag}} & \braket{\hat{a}^{\dag} \hat{a}^{\dag}} & \braket{\hat{a}^\dag \hat{a}} & \braket{\hat{a}^\dag \hat{b}^{\dag}}\\
  \braket{\hat{b}} & \braket{\hat{a}^{\dag} \hat{b}} & \braket{\hat{a}\hat{b}} & \braket{\hat{b}^\dag \hat{b}}
\end{vmatrix}$ & 4 & 2 & (1,2,3,5) \\

$D_\mathrm{VI}$ & $\begin{vmatrix}
1 & \braket{\hat{a}^\dag} & \braket{\hat{a}} & \braket{\hat{b}} & \braket{\hat{b}^\dag} \\
\braket{\hat{a}} & \braket{\hat{a}\hat{a}^\dag} & \braket{\hat{a}^2} & \braket{\hat{a}\hat{b}} & \braket{\hat{a}\hat{b}^\dag} \\
\braket{\hat{a}^\dag} & \braket{\hat{a}^{\dag 2} } & \braket{\hat{a}^\dag \hat{a}} & \braket{\hat{a}^\dag\hat{b}} & \braket{\hat{a}^\dag\hat{b}^\dag} \\
\braket{\hat{b}^\dag} & \braket{\hat{a}^\dag \hat{b}^\dag} & \braket{\hat{a}\hat{b}^\dag} & \braket{\hat{b}\hat{b}^\dag} & \braket{\hat{b}^{\dag 2}} \\
\braket{\hat{b}} & \braket{\hat{a}^\dag\hat{b}} & \braket{\hat{a}\hat{b}} & \braket{\hat{b}^2} & \braket{\hat{b}^\dag\hat{b}} \\
\end{vmatrix}$ & 5 & 2 & (1,2,3,4,5) \\

  $D_\mathrm{VII}$ &  $\begin{vmatrix}
  \braket{\hat{a}^{\dag} \hat{a}} & \braket{\hat{a}^{\dag}\hat{b}} & \braket{\hat{a}^{\dag} \hat{b}^{\dag}} \\
  \braket{\hat{a} \hat{b}^{\dag}} & \braket{\hat{b} \hat{b}^{\dag}} & \braket{\hat{b}^{\dag 2}} \\
  \braket{\hat{a} \hat{b}} & \braket{\hat{b}^2} & \braket{\hat{b}^\dag \hat{b}}
\end{vmatrix}$ & 3 & 2 & (3,4,5) \\

   $D_\mathrm{VIII}$ & $\begin{vmatrix}
    1 &   \braket{\hat{a}} &   \braket{\hat{b}} &   \braket{\hat{b}^{\dag}} \\
    \braket{\hat{a}^{\dag}} & \braket{\hat{a}^{\dag} \hat{a}} & \braket{\hat{a}^{\dag}\hat{b}} & \braket{\hat{a}^{\dag} \hat{b}^{\dag}} \\
    \braket{\hat{b}^{\dag}} & \braket{\hat{a} \hat{b}^{\dag}} & \braket{\hat{b} \hat{b}^{\dag}} & \braket{\hat{b}^{\dag 2}}\\
    \braket{\hat{b}} & \braket{\hat{a} \hat{b}} & \braket{\hat{b}^2} & \braket{\hat{b}^\dag \hat{b}}
\end{vmatrix}$ & 4 & 2 & (1,3,4,5) \\

\end{tabular}
\end{ruledtabular}
\label{tab:TMSV}
\end{table}

\begin{table}[h]
\caption{Suitable NPT criteria for the single-quantum-subtracted TMSV state $\ket{\Psi}_{\mathrm{SUB}}$. These determinants are plotted in Fig.~\ref{fig:tmsv_sub_results_fig}. The matrix $\mathbf{A}$ from which the determinant is calculated is provided, along with the dimension $d$, the order $n$ of the highest-order moment, and the rows/columns that are kept from the matrix $\mathbf{M}$.
Determinants $D_{\mathrm{I}}$ and $E_{\mathrm{I}-\mathrm{V}}$ are the only determinants with $d=2$ and $n\leq 4$ that exhibit negativity in the region of parameter space explored in Fig.~\ref{fig:tmsv_sub_results_fig}. By swapping $\hat{a}$ ($\hat{a}^\dag$) with $\hat{b}$ ($\hat{b}^\dag$), one may see that $E_{\mathrm{I}}=E_{\mathrm{IV}}$ and $E_{\mathrm{II}}=E_{\mathrm{V}}$.}
\begin{ruledtabular}
\begin{tabular}{l c c c r}
Determinant & Matrix form, $\mathrm{det}[\mathbf{A}]$  & Dimension, $d$ & Order, $n$ & Rows/Columns  \\
  \hline
  $D_{\mathrm{I}}$ & $\begin{vmatrix}
  \braket{\hat{a}^\dag \hat{a}} & \braket{\hat{a}^\dag \hat{b}^{\dag}}\\
  \braket{\hat{a}\hat{b}} & \braket{\hat{b}^\dag \hat{b}}
\end{vmatrix}$ & 2 & 2 & (3,5) \\
\vspace{0.2cm}
  $E_{\mathrm{I}}$ & $\begin{vmatrix}
      \braket{\hat{a}^\dag \hat{a} \hat{a}^\dag \hat{a}} & \braket{\hat{a}^\dag \hat{a} \hat{a}^\dag \hat{b}^\dag}\\
      \braket{\hat{a} \hat{a}^\dag \hat{a} \hat{b}} & \braket{\hat{a} \hat{a}^\dag \hat{b}^\dag \hat{b}}
  \end{vmatrix}$ & 2 & 4 & (7,12) \\
  \vspace{0.2cm}
  $E_{\mathrm{II}}$ & $\begin{vmatrix}
  \braket{\hat{a}^\dag \hat{a}^\dag \hat{a}\hat{a}} & \braket{\hat{a}^\dag \hat{a}^\dag \hat{a} \hat{b}^\dag}\\
  \braket{\hat{a}^\dag \hat{a} \hat{a} \hat{b}} & \braket{\hat{a}^\dag \hat{a} \hat{b}^\dag \hat{b}}
\end{vmatrix}$ & 2 & 4 & (8,13)  \\
\vspace{0.2cm}
  $E_{\mathrm{III}}$ & $\begin{vmatrix}
  \braket{\hat{a}^\dag \hat{a}^\dag \hat{a} \hat{a}} & \braket{\hat{a}^\dag \hat{a}^\dag \hat{b}^\dag \hat{b}^\dag} \\
  \braket{\hat{a}\hat{a}\hat{b}\hat{b}} & \braket{\hat{b}^\dag \hat{b}^\dag \hat{b}\hat{b}}
\end{vmatrix}$ & 2 & 4 & (8,15)  \\
\vspace{0.2cm}
  $E_{\mathrm{IV}}$ & $\begin{vmatrix}
  \braket{\hat{a}^\dag \hat{a} \hat{b} \hat{b}^{\dag}} & \braket{\hat{a}^\dag \hat{b}^\dag \hat{b} \hat{b}^\dag} \\
  \braket{\hat{a}\hat{b}\hat{b}^{\dag}\hat{b}} & \braket{\hat{b}^\dag \hat{b} \hat{b}^{\dag}\hat{b}}
\end{vmatrix}$ & 2 & 4 & (10,14)  \\

 $E_{\mathrm{V}}$ & $\begin{vmatrix}
  \braket{\hat{a}^\dag \hat{a} \hat{b}^{\dag} \hat{b}} & \braket{\hat{a}^\dag \hat{b}^\dag \hat{b}^\dag \hat{b}} \\
  \braket{\hat{a}\hat{b}^{\dag}\hat{b}\hat{b}} & \braket{\hat{b}^\dag \hat{b}^\dag \hat{b}\hat{b}}
\end{vmatrix}$ & 2 & 4 & (13,15)  \\
\end{tabular}
\end{ruledtabular}
\label{tab:TMSV_sub}
\end{table}

\begin{table}[h]
\caption{Suitable NPT criteria for the two-mode Schr{\"o}dinger-cat state $\ket{\Psi}_{\mathrm{CAT}}\propto \ket{\alpha}\ket{0}-\ket{0}\ket{\alpha}$. These determinants are plotted in Fig.~\ref{fig:results_TMSC}. The matrix $\mathbf{A}$ from which the determinant is calculated is provided, along with the dimension $d$, the order $n$ of the highest-order moment, and the rows/columns that are kept from the matrix $\mathbf{M}$.
Determinants $F_\mathrm{{I-VI}}$ are the only determinants with $d=2$ and $n\leq 4$, which exhibit negativity in the region of parameter space explored in Fig.~\ref{fig:results_TMSC}. Owing to the symmetry of the state $\ket{\Psi}_{\mathrm{CAT}}$, $F_{\mathrm{II}}=F_{\mathrm{V}}$ and $F_{\mathrm{III}}=F_{\mathrm{VI}}$.
We have also included $S_{\mathrm{III}}$ from Eq.~\eqref{eq:S3_matrix} for comparison.}
\begin{ruledtabular}
\begin{tabular}{l c c c r}
Determinant & Matrix form, $\mathrm{det}[\mathbf{A}]$ & Dimension, $d$ & Order, $n$ & Rows/Columns \\ 
  \hline
    \vspace{0.2cm}
    $F_{\mathrm{I}}$ & $\begin{vmatrix}
1 & \braket{\hat{a}\hat{b}^\dag}\\
\braket{\hat{a}^\dag \hat{b}} & \braket{\hat{a}^\dag \hat{a} \hat{b}^\dag \hat{b}}
\end{vmatrix}$ & 2 & 4 & (1,13) \\
\vspace{0.2cm}
  $F_{\mathrm{II}}$ &  $\begin{vmatrix}
      \braket{\hat{a} \hat{a}^\dag} & \braket{\hat{a}\hat{a} \hat{b}^\dag}\\
      \braket{\hat{a}^\dag \hat{a}^\dag \hat{b}} & \braket{\hat{a}^\dag \hat{a} \hat{b}^\dag \hat{b}}
  \end{vmatrix}$ & 2 & 4 & (2,13) \\
  \vspace{0.2cm}
  $F_{\mathrm{III}}$ & $\begin{vmatrix}
\braket{\hat{a}\hat{a}\hat{a}^\dag \hat{a}^\dag} & \braket{\hat{a}\hat{a}\hat{a}\hat{b}^\dag}\\
\braket{\hat{a}^\dag \hat{a}^\dag \hat{a}^\dag \hat{b}} & \braket{\hat{a}^\dag \hat{a} \hat{b}^\dag \hat{b}}
\end{vmatrix}$  & 2 & 4 & (6,13) \\
  \vspace{0.2cm}
  $F_{\mathrm{IV}}$ & $\begin{vmatrix}
\braket{\hat{a}\hat{a}^\dag \hat{b}\hat{b}^\dag} & \braket{\hat{a}\hat{a}\hat{b}^\dag \hat{b}^\dag}\\
\braket{\hat{a}^\dag \hat{a}^\dag \hat{b}\hat{b}} & \braket{\hat{a}^\dag \hat{a} \hat{b}^\dag \hat{b}}
\end{vmatrix}$ & 2 & 4 & (9, 13) \\
  \vspace{0.2cm}
    $F_{\mathrm{V}}$ & $\begin{vmatrix}
      \braket{\hat{b} \hat{b}^\dag} & \braket{\hat{a} \hat{b}^\dag \hat{b}^\dag}\\
      \braket{\hat{a}^\dag \hat{b} \hat{b}} & \braket{\hat{a}^\dag \hat{a} \hat{b}^\dag \hat{b}}
  \end{vmatrix}$ & 2 & 4 & (4, 13) \\
  \vspace{0.2cm}
      $F_{\mathrm{VI}}$ & $\begin{vmatrix}
\braket{\hat{b}\hat{b}\hat{b}^\dag \hat{b}^\dag} & \braket{\hat{a}\hat{b}^\dag \hat{b}^\dag \hat{b}^\dag}\\
\braket{\hat{a}^\dag \hat{b} \hat{b} \hat{b}} & \braket{\hat{a}^\dag \hat{a} \hat{b}^\dag \hat{b}}
\end{vmatrix}$ & 2 & 4 & (11, 13) \\
  \vspace{0.2cm}
  $S_{\mathrm{III}}$ & $\begin{vmatrix}
1 & \braket{\hat{b}^\dag} & \braket{\hat{a}\hat{b}^\dag} \\
\braket{\hat{b}} & \braket{\hat{b}^\dag \hat{b}} & \braket{\hat{a} \hat{b}^\dag \hat{b}}\\
\braket{\hat{a}^\dag \hat{b}} & \braket{\hat{a}^\dag \hat{b}^\dag \hat{b}} & \braket{\hat{a}^\dag \hat{a} \hat{b}^\dag \hat{b}}
\end{vmatrix}$ & 3 & 4 & (1, 5, 13) \\
\end{tabular}
\label{tab:TMSC}
\end{ruledtabular}
\end{table}

\section{Invariance of NPT criteria under local rotations\label{app:invariance}}

A bipartite state $\hat{\rho}_\mathcal{AB}$, comprised of two subsystems $\mathcal{A}$ and $\mathcal{B}$, is mapped onto the state ${\hat{\tilde{\rho}}}_\mathcal{AB}={\hat{U}}(\theta_A,\theta_B)\hat{\rho}_\mathcal{AB}{\hat{U}}^\dag(\theta_A,\theta_B)$ where ${\hat{U}}(\theta_A,\theta_B)=\exp (-i\theta_{A}\hat{a}^\dag\hat{a})\exp(-i\theta_B \hat{b}^\dag \hat{b})$. Considering a specific matrix of moments $\mathbf{A}$ from which we will construct the entanglement test $\mathrm{det}[\mathbf{A}]<0$ for $\hat{\rho}_\mathcal{AB}$, 
the matrix element with indices $i$ and $j$ is the moment
${A}_{ij}=\braket{\hat{a}^{\dag {q_i}}\hat{a}^{p_i} \hat{a}^{\dag n_j}\hat{a}^{m_j} \hat{b}^{\dag l_j}\hat{b}^{k_j}\hat{b}^{\dag r_i}\hat{b}^{s_i}}$. Calculating the same moments for ${\hat{\tilde{\rho}}}_\mathcal{AB}$ will yield a different matrix of moments denoted by $\tilde{\mathbf{A}}$. 
For this rotated state, the moment corresponding to the indices $i$ and $j$ is then
$
\tilde{A}_{ij}=\langle{{\hat{U}}^\dag(\theta_A,\theta_B)a^{\dag {q_i}}a^{p_i} a^{\dag n_j}a^{m_j} b^{\dag l_j}b^{k_j}b^{\dag r_i}b^{s_i}{\hat{U}}(\theta_A,\theta_B)\rangle}$.

Using the identity $\exp(i\theta\hat{a}^\dag \hat{a}) \hat{a} \exp(-i\theta\hat{a}^\dag \hat{a})=\hat{a} e^{-i\theta}$, and similarly for $\hat{b}$, it follows that
\begin{equation}
\begin{split}
\tilde{A}_{ij}&={A}_{ij}\mathrm{e}^{i\theta_{A}(q_i-p_i+n_j-m_j)+i\theta_B(l_j-k_j+r_i-s_i)}~.
\end{split}
\end{equation}
As discussed in Section~\ref{sec:overview_entanglement}, the two sets of multi-indices have the same ordering such that the $N^{\mathrm{th}}$ multi-index is the same for both $(p_N,q_N,r_N,s_N)=(n_N,m_N,k_N,l_N)$. From this property we can write the indices in the following manner
\begin{equation}
\begin{split}
\tilde{A}_{ij}&={A}_{ij}\mathrm{e}^{i\theta_{A}(q_i-p_i)+i\theta_B(r_i-s_i)}\mathrm{e}^{-i\theta_A(q_j-p_j)-i\theta_B(r_j-s_j)}~,
\end{split}
\end{equation}
and using the substitution $\phi_{u}=\theta_{A}(q_u-p_u)+\theta_B(r_u-s_u)$ where $u=\{i,j\}$, we then have
\begin{equation}
\tilde{A}_{ij}={A}_{ij}e^{i\phi_i-i\phi_j}~.
\end{equation}
Hence, matrix $\mathbf{A}$ is mapped onto $\tilde{\mathbf{A}}$ by the following transformation $\tilde{\mathbf{A}}=\mathbf{{U}}^\dag\mathbf{A}\mathbf{\hat{U}}$ where $\mathbf{{U}}$ is a diagonal, unitary matrix whose $i^{\mathrm{th}}$ and $j^{\mathrm{th}}$ elements are ${U}_{ij}=e^{-i\phi_i}\delta_{ij}$. Since the determinant of a unitary matrix is 1, it is evident that $\mathrm{det}[\tilde{\mathbf{A}}]=\mathrm{det}[\mathbf{A}]$. Therefore, the NPT criterion $\mathrm{det}[\mathbf{A}]<0$ is invariant as a state undergoes local rotations. In general, however, this result does not hold for other local operations such as displacements. 

\section{Unbiased determinant estimation and error analysis}\label{app:error_prop}

Suppose we have a scalar-valued function $f(\textbf{x})=f(x_1,x_2,\ldots,x_n)$ where $\textbf{x}$ represents a vector of independent sample means to be measured. Each element $x_i$ has mean $\mu_{i}=E[x_{i}]$ and variance $\sigma^2_{i}/M_i=E\left[(x_i-\mu_i)^2\right]$ where $M_i$ independent measurements are assigned to each $x_i$. 
In this work, we are interested in estimating the determinant of the Hermitian $d\times d$ submatrix $\textbf{A}$.
We denote the matrix of sample means of the matrix elements $A_{ij}$ as $\bar{\mathbf{A}}$. For our purposes the function $f(\mathbf{x})$ plays the role of  an unbiased estimator of $\mathrm{det}[\textbf{A}]$, that is, $E[f(\mathbf{x})]=\mathrm{det}[\mathbf{A}]$, where $\mathbf{x}$ comprises all sample means $\bar{B}_{ij,p}$ required to measure the matrix elements $\bar{A}_{ij}=\bar{B}_{ij,0}+i\bar{B}_{ij,1}\left(1-\delta_{ij}\right)$. 
The matrix $E[\bar{\mathbf{A}}] = \mathbf{A}$ formed of the true values is Hermitian, however, for any one experimental realization of submatrix $\bar{\mathbf{A}}$, we cannot ensure Hermiticity, and hence $\mathrm{det}[\bar{\textbf{A}}]$ will not be real in general. To ensure reality, we define our estimator $f(\mathbf{x})=\mathrm{Re}\{\mathrm{det}[\bar{\textbf{A}}]\}$.

We make use of the totally antisymmetric Levi--Civita tensor $\varepsilon_{i_1i_2...i_d}$ in $d$ dimensions to write
\begin{align}\label{eq:appendix_det_A}
    \mathrm{Re}\{\mathrm{det}[\bar{\textbf{A}}]\}=\sum_{j_1,j_2,...,j_d}^d\varepsilon_{j_1j_2...j_d}\mathrm{Re}\{\bar{A}_{1j_1}\bar{A}_{2j_2}...\bar{A}_{dj_d}\} \,.
\end{align}
We notice that each product in the sum in Eq.~\eqref{eq:appendix_det_A} is comprised of statistically independent $\bar{A}_{ij}$. Hence, $E[\mathrm{Re}\{\mathrm{det}[\bar{\textbf{A}}]\}] = \mathrm{Re}\{\mathrm{det}[E[\bar{\mathbf{A}}]]\} = \mathrm{det}[\textbf{A}]$, so $f(\mathbf{x})$ is unbiased. 
The variance of the function $f(\textbf{x})$ is
\begin{align}\label{eq:appendix_var_f_x}
    \mathrm{Var}(f(\mathbf{x}))&=E[(f(\mathbf{x})-E[f(\mathbf{x})])^2] \, .
\end{align}
We expand $f(\mathbf{x})$ in a Taylor series about the population means $\mu_i$ as 
\begin{align}\label{eq:appendix_f_x}
f(\mathbf{x})= f(\mathbf{\mu})+\sum_i^d \frac{\partial f}{\partial x_i}\Bigg|_{{\mu_i}}(x_i-\mu_i)+\frac{1}{2!}\sum_{i,j}^d \frac{\partial^2 f}{\partial x_i \partial x_j}\Bigg|_{{\mu_i}}(x_i-\mu_i)(x_j-\mu_j)+\mathcal{O}\left((x-\mu)^3\right)\, ,
\end{align}
where we have omitted the indices in the last term to indicate cubic and higher order terms.
For $f(\mathbf{x}) = \mathrm{Re}\{\mathrm{det}[\bar{\textbf{A}}]\} $,
\begin{align}\label{eq:derivatives}
\frac{\partial\,\mathrm{Re}\{\mathrm{det}[\bar{\textbf{A}}]\}}{\partial \bar{B}_{ij,p}}\Bigg|_{\langle{\hat{B}_{ij,p}}\rangle} &= \frac{1}{(d-1)!}\sum_{\substack{i_2, ..., i_d\\ j_2, ..., j_d}}\varepsilon_{i, i_2, ..., i_d}\varepsilon_{j, j_2, ..., j_d}\mathrm{Re}\{(i)^p{A}_{i_2j_2}...A_{i_dj_d}\} =\mathrm{Re}\{(-i)^p\mathrm{adj}[\mathbf{A}]_{ij}\}\, ,\\\nonumber
\frac{\partial^2 \mathrm{Re}\{\mathrm{det}[\bar{\textbf{A}}]\}}{\partial \bar{B}_{ij,p}\partial \bar{B}_{kl,q}}\Bigg|_{\langle{\hat{B}_{ij,p}}\rangle} &= \frac{1}{(d-2)!}\sum_{\substack{i_3, i_4, ..., i_d \\ j_3, j_4, ..., j_d}}\varepsilon_{iki_3...i_d}\varepsilon_{jlj_3...j_d}\mathrm{Re}\{(i)^{p+q}A_{i_3j_3}...A_{i_dj_d}\} \,,
\end{align}
where $\mathrm{adj}[\mathbf{A}]$ is the adjugate of the matrix $\mathbf{A}$.
Using these, we can write $\mathrm{Var}(\mathrm{Re}\{\mathrm{det}[\bar{\textbf{A}}]\}) = (\Delta\mathrm{Re}\{\mathrm{det}[\bar{\textbf{A}}]\})^2$ as 
\begin{align}\label{eq:delta_detA2_secorder}
(\Delta\mathrm{Re}\{\mathrm{det}[\bar{\textbf{A}}]\})^2 &= \sum_{i,j,p}\left(\frac{\partial\,\mathrm{Re}\{\mathrm{det}[\bar{\textbf{A}}]\}}{\partial \bar{B}_{ij,p}}\right)^2\Bigg|_{\langle{\hat{B}_{ij,p}}\rangle}\frac{\mathrm{Var}(\hat{B}_{ij,p})}{M_{ij,p}}\\\nonumber
&+ \frac{1}{2}\sum_{\substack{i,j,k,l,p,q\\ i\neq k,j\neq l}}\left(\frac{\partial^2 \mathrm{Re}\{\mathrm{det}[\bar{\textbf{A}}]\}}{\partial \bar{B}_{ij,p}\partial \bar{B}_{kl,q}}\right)^2\Bigg|_{\langle{\hat{B}_{ij,p}}\rangle}\frac{\mathrm{Var}(\hat{B}_{ij,p})}{M_{ij,p}}\frac{\mathrm{Var}(\hat{B}_{kl,q})}{M_{kl,q}}+\mathcal{O}\left(\left(\frac{\mathrm{Var}(\hat{B})}{M}\right)^3\right)\, ,
\end{align}
which requires $M_{ij,p}$ to be sufficiently large such that $\mathrm{Var}(\bar{B}_{ij,p}) =\mathrm{Var}(\hat{B}_{ij,p})/M_{ij,p}\ll1$ such that cubic and higher order terms can be ignored, and we have again neglected the indices on the cubic term.
In the second order term, we have ignored cross terms with coefficients of the form
$(\partial^2 \mathrm{Re}\{\mathrm{det}[\bar{\textbf{A}}]\}/\partial \bar{B}_{i_1j_1,p_1}\partial \bar{B}_{k_1l_1,q_1})\cdot(\partial^2 \mathrm{Re}\{\mathrm{det}[\bar{\textbf{A}}]\}/\partial \bar{B}_{i_2j_2,p_2}\partial \bar{B}_{k_2l_2,q_2})$ for $(i_1,j_1,k_1,l_1;p_1,q_1)\neq(i_2,j_2,k_2,l_2;p_2,q_2)$. These terms vanish due to the presence of terms of the form $E[(\bar{B}_{i_1j_1,p_1}-\langle{\hat{B}_{i_1j_1,p_1}}\rangle)(\bar{B}_{k_1l_1,q_1}-\langle{\hat{B}_{k_1l_1,q_1}}\rangle)(\bar{B}_{i_2j_2,p_2}-\langle{\hat{B}_{i_2j_2,p_2}}\rangle)(\bar{B}_{k_2l_2,q_2}-\langle{\hat{B}_{k_2l_2,q_2}}\rangle)]$ with $i_1\neq k_1$, $j_1\neq l_1$, $i_2\neq k_2$, $j_2\neq l_2$ since the $\varepsilon$ symbol vanishes for two repeated indices.

The quantity $(\Delta{\mathrm{Re}\{\mathrm{det}[\bar{\mathbf{A}}]})^2$ can be used to estimate the width of the determinant distribution. We optimally assign measurement numbers $M_{ij,p}$ by minimizing this quantity under the constraint of a total measurement number $M_{\text{tot}}$. The full determinant distribution is used to calculate the statistical power of the entanglement hypothesis test given as the probability of the determinant being negative when the true value $\mathrm{det}[\mathbf{A}]<0$.

\section{Optimal measurement allocation\label{app:minimizing_std}}

The error up to second order on the determinant estimator $\mathrm{Re}\{\mathrm{det}[\bar{\mathbf{A}}]\}$ is given by Eq.~\eqref{eq:delta_detA2_secorder} by inserting the expressions from Eq.~\eqref{eq:derivatives}. Now, we wish to minimize the expression for $(\Delta \mathrm{Re}\{\mathrm{det}[\bar{\mathbf{A}}]\})^2$ under the constraint of a fixed total number of measurements given by
\begin{align}
M_{\mathrm{tot}}=\sum_{i}M_{ii,0}+2\sum_{i<j,p}M_{ij,p} \,.
\end{align}
Using the method of Lagrange multipliers, we define a new function
\begin{align}
    \mathcal{L}(\{M_{ij,p}\},\lambda)= (\Delta \mathrm{Re}\{\mathrm{det}[\bar{\mathbf{A}}]\})^2+\lambda\left(\sum_{i}M_{ii,0}+2\sum_{i<j,p}M_{ij,p}-M_{\mathrm{tot}}\right)\,,
\end{align}
where $\lambda$ is a Lagrange multiplier for the $M_{\mathrm{tot}}$ constraint. Differentiating $\mathcal{L}(\{M_{ij,p}\},\lambda)$ with respect to the $M_{ij,p}$ gives
\begin{align}\label{eq:optimization_eqs}
    \frac{\partial \mathcal{L}}{\partial M_{jj,0}} &= -(\mathrm{adj}[\mathbf{A}]_{jj})^2\frac{\mathrm{Var}(\hat{B}_{jj,0})}{M_{jj,0}^2}+\lambda-\sum_{k,l,q}\left(\frac{\partial^2 \mathrm{Re}\{\mathrm{det}[\bar{\textbf{A}}]\}}{\partial \bar{B}_{jj,p}\partial \bar{B}_{kl,q}}\right)^2\Bigg|_{\langle{\hat{\mathbf{B}}}\rangle}\frac{\mathrm{Var}(\hat{B}_{jj,p})}{M_{jj,p}^2}\frac{\mathrm{Var}(\hat{B}_{kl,q})}{M_{kl,q}}\, , \\\nonumber
    \frac{\partial \mathcal{L}}{\partial M_{ij,0}}&=-2(\mathrm{Re}[\mathrm{adj[\mathbf{A}}]]_{ij})^2\frac{\mathrm{Var}(\hat{B}_{ij,0})}{M_{ij,0}^2}+2\lambda-2\sum_{k,l,q}\left(\frac{\partial^2 \mathrm{Re}\{\mathrm{det}[\bar{\textbf{A}}]\}}{\partial \bar{B}_{ij,0}\partial \bar{B}_{kl,q}}\right)^2\Bigg|_{\langle{\hat{\mathbf{B}}}\rangle}\frac{\mathrm{Var}(\hat{B}_{ij,0})}{M_{ij,1}^2}\frac{\mathrm{Var}(\hat{B}_{kl,q})}{M_{kl,q}}\, , \\\nonumber
    \frac{\partial \mathcal{L}}{\partial M_{ij,1}}&=-2(\mathrm{Im}[\mathrm{adj[\mathbf{A}}]]_{ij})^2\frac{\mathrm{Var}(\hat{B}_{ij,1})}{M_{ij,1}^2}+2\lambda-2\sum_{k,l,q}\left(\frac{\partial^2 \mathrm{Re}\{\mathrm{det}[\bar{\textbf{A}}]\}}{\partial \bar{B}_{ij,1}\partial \bar{B}_{kl,q}}\right)^2\Bigg|_{\langle{\hat{\mathbf{B}}}\rangle}\frac{\mathrm{Var}(\hat{B}_{ij,1})}{M_{ij,1}^2}\frac{\mathrm{Var}(\hat{B}_{kl,q})}{M_{kl,q}}\, .
\end{align}
Setting the LHS of each equation in Eq.~\eqref{eq:optimization_eqs} to zero, we get a system of non-linear equations that in general cannot be solved analytically for $M_{ij,p}$, and is solved numerically with \textit{NPyT}. 

\subsection{Analytical allocation for dimension 2}
We now illustrate this optimization scheme analytically for $d=2$. We consider the matrix $\bar{\mathbf{A}}$ of the form
\begin{equation}\label{eq:2x2eg}
    \bar{\mathbf{A}}=\begin{pmatrix}
        \bar{a}&\bar{c_1}+i\bar{d_1}\\
        \bar{c_2}+i\bar{d_2}&\bar{b}
    \end{pmatrix}\, ,
\end{equation}
where, $\bar{a},\bar{b},\bar{c_1},\bar{c_2},\bar{d_1},\bar{d_2}$ are all real random variables satisfying $E[\bar{a}]=\mu_a$, $E[\bar{b}]=\mu_b$, $E[\bar{c_1}]=E[\bar{c_2}]=\mu_c$, $E[\bar{d_1}]=-E[\bar{d_2}]=\mu_d$, $\mathrm{Var}(\bar{a})=\sigma^2_a/M_a$, $\mathrm{Var}(\bar{b})=\sigma^2_b/M_b$, $\mathrm{Var}(\bar{c}_1)=\mathrm{Var}(\bar{c}_2)=\sigma^2_c/M_c$, $\mathrm{Var}(\bar{d}_1)=\mathrm{Var}(\bar{d}_2)=\sigma^2_d/M_d$. Inputting these into Eq.\eqref{eq:optimization_eqs} and setting the LHS to zero, we get the four equations for the four unknowns,
\begin{align}\label{eq:2x2_optimization}
    M_{a}&:\quad -\mu_b^2\frac{\sigma^2_a}{M_a^2}+\lambda-\frac{\sigma^2_a\sigma^2_b}{M_a^2M_b}=0\, ,\nonumber\\
    M_{b}&:\quad -\mu_a^2\frac{\sigma^2_b}{M_b^2}+\lambda-\frac{\sigma^2_a\sigma^2_b}{M_aM_b^2}=0\, ,\\
    M_{c}&:\quad -\mu_c^2\frac{\sigma_c^2}{M_c^2}+\lambda-\frac{\sigma_c^4}{M_c^3}=0\nonumber\, ,\\
    M_{d}&:\quad -\mu_d^2\frac{\sigma_d^2}{M_d^2}+\lambda-\frac{\sigma_d^4}{M_d^3}=0\, .\nonumber   
\end{align}
For simplicity, we assume $\mu_d=0$ and the rest non-zero. We also assume that for the $M_a,M_b,M_c$ equations in Eq.~\eqref{eq:2x2_optimization}, we can ignore the terms that are cubic in $M_{a-c}$ compared to the quadratic terms. This is a good approximation when $\sigma^2_a/M_a, \sigma^2_b/M_b, \sigma^2_c/M_c$ are small. In this approximation, we can write down analytical solutions for $M_a,M_b,M_c,M_d$ as
\begin{align}
    M_a^{\mathrm{opt}} &\approx \mu_b\frac{\sigma_a}{\sqrt{\lambda}}\, ,\nonumber\\
    M_b^{\mathrm{opt}}&\approx\mu_a\frac{\sigma_b}{\sqrt{\lambda}}\, ,\\
    M_c^{\mathrm{opt}}&\approx\mu_c\frac{\sigma_c}{\sqrt{\lambda}}\, ,\nonumber\\
    M_d^{\mathrm{opt}}&=\frac{\sigma_d^{\frac{4}{3}}}{\lambda^{\frac{1}{3}}} \,.\nonumber
\end{align}

\section{Monte Carlo method for the determinant distribution}\label{app:monte_carlo}
In \textit{NPyT}, the determinant distribution is obtained numerically by sampling each estimator $\bar{B}_{ij,p}$ either from a Gaussian or a gamma distribution depending on the properties of the corresponding operator, and then generating a histogram. The Hermitian operators (those $\hat{O}_{ij}$ obeying $\hat{O}_{ij} = \hat{O}^{\dag}_{ij}$) are positive-semidefinite and are given as a product of annihilation and creation operators of the two modes $a$ and $b$. Hence, to approximate the distribution of the estimators of their quantum expectation values we use gamma distributions with means $\langle \hat{O}_{ij}\rangle$ and variances $\textrm{Var}(\hat{O}_{ij})/M_{ij}$ for the optimal number of measurements $M_{ij}$. Although the positively skewed distribution of the estimator of $\langle\hat{O}_{ij}\rangle$ can be approximated by a narrow Gaussian centered at the non-negative mean value, this would require a large number of measurements. Using a gamma distribution captures the nonnegativity, finite-sample skewness, and reduces to the Gaussian limit for large measurement number. We further note that the main diagonal elements of $\mathbf{A}$ are always Hermitian. The shape and scale of the gamma distributions are determined uniquely in terms of the optimal measurements $M_{ij}$, $\langle\hat{O}_{ij}\rangle$ and $\textrm{Var}(\hat{O}_{ij})$ given by $M_{ij}\cdot\langle\hat{O}_{ij}\rangle^2/\textrm{Var}(\hat{O}_{ij})$ and $\textrm{Var}(\hat{O}_{ij})/(M_{ij}\cdot \langle\hat{O}_{ij}\rangle)$ respectively. With this information, we are able to match the means and variances of the gamma distributions with those of the estimators.

As for the non-Hermitian operators in $\mathbf{A}$, we approximate the distributions of the estimators of their real and imaginary parts $\bar{B}_{ij,0}$, $\bar{B}_{ij,1}$ as Gaussian distributions with means $\langle\hat{B}_{ij,p}\rangle$ and variances $\mathrm{Var}(\hat{B}_{ij,p})/M_{ij,p}$. We further numerically compute their excess kurtoses to establish when this approximation is valid. The excess kurtoses of $\bar{B}_{ij,p}$ are given by $(\langle{(\hat{B}_{{ij,p}}-\langle{\hat{B}_{ij,p}}\rangle})^4\rangle-3\cdot\mathrm{Var}(\hat{B}_{ij,p}))/M_{ij,p}$. In \textit{NPyT}, we set an excess kurtosis threshold and flag the instances (determined by the state, NPT criterion $\mathbf{A}$, and the set of optimal measurements) when the absolute value of the excess kurtosis of any of the $\bar{B}_{ij,p}$ exceeds this threshold. $M_{\mathrm{tot}}$ must be increased in order for the flagged $\bar{B}_{ij,p}$ to obey the threshold.

\section{Open-system dynamics}
\label{app:open_system}
All the moments required to calculate $\mathrm{det}[\mathbf{A}]$ and $\Delta\mathrm{Re}\{\mathrm{det}[\bar{\mathbf{A}}]\}$ are in the form $\braket{\hat{a}^{\dag m}\hat{a}^{n}\hat{b}^{\dag k}\hat{b}^{l}}$. From Eqs~\eqref{eq:beam_splitter}, \eqref{eq:environment_corr_1}, and \eqref{eq:environment_corr_2}, it can be seen that these expectation values, which have been subject to open-system dynamics, depend only on the expectation values with respect to the initial quantum state $\langle{\hat{a}^{\dag m}_0\hat{a}^{ n}_0\hat{b}^{\dag k}_0\hat{b}^{ l}_0\rangle}$ and the environmental parameters $\eta$ and $\bar{N}$. 

The result of Eq.~\eqref{eq:environment_correlations} may be derived using the Isserlis--Wick theorem as follows
\begin{eqnarray}
\braket{\hat{a}^{\dag p}_E \hat{a}_E^q}&=&\delta_{pq}p! \braket{\hat{a}^{\dag}_E \hat{a}_E}^p\nonumber\\
&=&\delta_{pq}p! \bar{N}^p.
\end{eqnarray}
Note that when applying Eq.~\eqref{eq:environment_correlations} to calculate the moments $\braket{\hat{a}^{\dag m}\hat{a}^{n}\hat{b}^{\dag k}\hat{b}^{l}}$, we assume that the noise operators of different modes, i.e. $\hat a_E$ and $\hat b_E$, are independent.

\subsection{Equivalence of beam-splitter-model and Heisenberg--Langevin-equation approaches}
\label{app:beamsplitter}

Here, we demonstrate the equivalence between the beam-splitter model for loss, outlined above and in the main text, with the approach that utilizes the Heisenberg--Langevin equations. These Heisenberg--Langevin equations for an open system are given by
\begin{equation}
\label{eq:Heisenberg-Lang}
    \dot{\hat{\mathbf{a}}}(t)=\frac{i}{\hbar}[\hat{H},\hat{\mathbf{a}}(t)]-\kappa \hat{\mathbf{a}}(t)+\sqrt{2\kappa}\hat{\mathbf{a}}_{\mathrm{in}}(t)~,
\end{equation}
where $\hat{\mathbf{a}}=(\hat{a}(t), \hat{a}^\dag(t))^{\mathrm{T}}$, the input field operators are $\hat{\mathbf{a}}_{\mathrm{in}}=(\hat{a}_{\mathrm{in}}(t),\hat{a}^\dag_{\mathrm{in}}(t))^{\mathrm{T}}$, and $\kappa$ is the amplitude decay rate. The input operators obey the following relations:
\begin{subequations}
\begin{align}
\langle{\hat{\mathbf{a}}_{\mathrm{in}}(t)\rangle}&=0,\\
    \langle{\hat{a}^\dag_{\mathrm{in}}(t)\hat{a}_{\mathrm{in}}(t')\rangle}&=\bar{N}\delta(t-t'),\\
    \langle{\hat{a}_{\mathrm{in}}(t)\hat{a}^\dag_{\mathrm{in}}(t')\rangle}&=(\bar{N}+1)\delta(t-t')~,
    \end{align}
\end{subequations}
where $\bar{N}$ is the mean thermal occupation number of the bath.
%
In the interaction picture, the solution to Eq.~\eqref{eq:Heisenberg-Lang} is
\begin{equation}
\label{eq:evolution}
\begin{split}
    \hat{\mathbf{a}}(t)&=\hat{\mathbf{a}}(0)\mathrm{e}^{-\kappa t}+\sqrt{2\kappa}\int_{0}^{t}dt'\hat{\mathbf{a}}_{\mathrm{in}}(t')\mathrm{e}^{-\kappa(t-t')}~,
\end{split}
\end{equation}
where $t$ is the time over which the system interacts with the environment. Note Eq.~\eqref{eq:Heisenberg-Lang} can be also applied to the second mode by substituting $\mathbf{a}$ for $\mathbf{b}$. In doing so, we assume both subsystems have the same decay rate $\kappa$. However, our approach can be generalized to subsystems with different amplitude decay rates. Throughout, we also assume that the noise operators acting on each subsystem are independent.

From the time-dependent matrix elements ${A}_{ij}(t)=\braket{\hat{B}_{ij,0}(t)}+i\braket{\hat{B}_{ij,1}(t)}$, one may calculate $\mathrm{det}[\mathbf{A}(t)]$ by inserting the Heisenberg operators in Eq.~\eqref{eq:evolution} into the quantum expectation values $\braket{\hat{a}^{\dag m}(t)\hat{a}^{n}(t)\hat{b}^{\dag k}(t) \hat{b}^{l}(t)}$. Similarly, to find $\Delta\mathrm{Re}\{\mathrm{det}[\bar{\mathbf{A}}]\}$ we can insert the time-dependent solutions of Eq.~\eqref{eq:evolution} into Eq.~\eqref{eq:delta_detA2_secorder_main}, replacing $\mathbf{A}$ and $\mathrm{Var}(\hat{B}_{ij,p})$ with $\mathbf{A}(t)$ and $\mathrm{Var}(\hat{B}_{ij,p}(t))$, respectively.

To show the equivalence between the beam-splitter-model and Heisenberg--Langevin-equation approaches, we compare Eq.~\eqref{eq:evolution} to Eq.~\eqref{eq:beam_splitter}. Firstly, comparing the first term in each equation we can see a direct equivalence by setting $\sqrt{\eta}$ equal to $\mathrm{e}^{-\kappa t}$ and $\hat{a}(0)$ to $\hat{a}_{0}$, as noted in the main text. Equivalence between the second terms, describing the environmental interactions may be seen by comparing the term $\sqrt{1-\eta}\hat{a}_{E}$ with the noise term
\begin{equation*}
    \sqrt{2\kappa}\int_{0}^{t}dt'{\hat{a}}_{\mathrm{in}}(t')\mathrm{e}^{-\kappa(t-t')}.
\end{equation*}
However, in this work, we are interested in the expectation values of moments of field operators. From Eq.~\eqref{eq:environment_correlations}, the environmental terms in the beam-splitter model for loss give the correlations
\begin{equation}\label{eq:comparing_correlations_BS}
    (1-\eta)^{(p+q)/2}\braket{\hat{a}^{\dag p}_E \hat{a}_E^q}=\delta_{pq}p!(1-\eta)^p\bar{N}^p.
\end{equation}
(Note, we remind the reader that the environmental modes are uncorrelated with the {initial} field operators of the entangled state and so all cross-correlations between the environment and the initial state are zero). Now, the environmental correlations in the Heisenberg--Langevin approach may be computed using the Isserlis--Wick theorem as follows
\begin{equation}
\label{eq:comparing_correlations_HL}
    \begin{split}
    &\langle{\bigg[\sqrt{2\kappa}\int_{0}^t dt' \hat{a}_{\mathrm{in}}^\dag(t')\mathrm{e}^{-\kappa(t-t')}\bigg]^{p}}
    \bigg[\sqrt{2\kappa}\int_{0}^t dt' \hat{a}_{\mathrm{in}}(t')\mathrm{e}^{-\kappa(t-t')}\bigg]^{q}\rangle \\
    &~~~~~~~=(2\kappa)^{\frac{p+q}{2}}\int_{0}^{t}\ldots \int_{0}^{t}dt^{(1)}_1\ldots dt^{(1)}_p dt^{(2)}_1\ldots dt^{(2)}_q  \mathrm{e}^{-\kappa(t-t^{(1)}_1)}\ldots \mathrm{e}^{-\kappa(t-t^{(1)}_p)} \mathrm{e}^{-\kappa(t-t^{(2)}_1)}\ldots\mathrm{e}^{-\kappa(t-t^{(2)}_q)}\\
    &~~~~~~~~~~~~~\langle \hat{a}^{\dag}_{\mathrm{in}}(t^{(1)}_1)\ldots \hat{a}^{\dag}_{\mathrm{in}}(t^{(1)}_p)\hat{a}_{\mathrm{in}}(t^{(2)}_1)\ldots \hat{a}_{\mathrm{in}}(t^{(2)}_q)\rangle\\
    &~~~~~~~=\delta_{pq} (2\kappa)^{\frac{p+q}{2}}p!\bigg[\int_{0}^{t}\int_{0}^{t}dt' dt'' \mathrm{e}^{-\kappa(t-t')}\mathrm{e}^{-\kappa(t-t'')}\langle \hat{a}_{\mathrm{in}}^\dag(t')\hat{a}_{\mathrm{in}}(t'')\rangle \bigg]^p \\
    &~~~~~~~=\delta_{pq} (2\kappa)^{p}p!\bigg[\int_{0}^{t}\int_{0}^{t}dt' dt'' \mathrm{e}^{-\kappa(t-t')}\mathrm{e}^{-\kappa(t-t'')}\bar{N}\delta(t-t') \bigg]^p \\
    &~~~~~~~=\delta_{pq} (2\kappa)^{p}p!\bigg[\int_{0}^{t}dt' \mathrm{e}^{-2\kappa(t-t')}\bar{N} \bigg]^p\\
    &~~~~~~~=\delta_{pq} (2\kappa)^{p}p! \bigg[\frac{1-\mathrm{e}^{-2\kappa t}}{2\kappa}\bigg]^p \bar{N}^p\\
    &~~~~~~~=\delta_{pq}p!( 1-\mathrm{e}^{-2\kappa t})^p\bar{N}^p.
    \end{split}
\end{equation}
Thus, comparison between the environmental correlations of Eqs~\eqref{eq:comparing_correlations_BS} and \eqref{eq:comparing_correlations_HL} shows that the beam-splitter-model and Heisenberg--Langevin approaches are indeed equivalent when $\sqrt{\eta}=\mathrm{e}^{-\kappa t}$.

\end{widetext}

\end{document}